\documentclass[sigconf, nonacm, pdfa]{acmart}

\newcommand\vldbdoi{10.14778/3746405.3746406}
\newcommand\vldbpages{2761 - 2774}
\newcommand\vldbvolume{18}
\newcommand\vldbissue{9}
\newcommand\vldbyear{2025}
\newcommand\vldbauthors{\authors}
\newcommand\vldbtitle{\shorttitle} 
\newcommand\vldbavailabilityurl{https://github.com/iDC-NEU/CCaaS}
\newcommand\vldbpagestyle{empty}

\usepackage{enumerate}
\usepackage[linesnumbered, ruled, vlined, noend, resetcount]{algorithm2e} 
\usepackage{graphics}
\usepackage{balance}
\usepackage{graphicx}
\usepackage{subfig}
\usepackage{amsthm}
\usepackage{bm}
\usepackage{algpseudocode}
\usepackage{listings}
\usepackage{enumitem}
\usepackage{multirow}
\usepackage{float}
\usepackage{url}
\usepackage[normalem]{ulem}
\usepackage{comment}
\usepackage{amsmath}
\usepackage{epsfig}
\usepackage{booktabs}
\usepackage{lipsum}
\usepackage{array}
\usepackage{bbding} 
\usepackage{setspace}
\usepackage{listings}

\usepackage[utf8]{inputenc}
\usepackage[UKenglish]{babel}
\usepackage{colorprofiles}
\usepackage[a-2b, mathxmp]{pdfx}[2018/12/22]
\hypersetup{pdfstartview=}

\newcommand{\Paragraph} [1] {\smallskip\noindent{\bf #1. }}

\newcommand{\oursys}{\texttt{CCaaS}\xspace}

\newcommand{\ie}{\emph{i.e.,}\xspace}
\newcommand{\eg}{\emph{e.g.,}\xspace}

\newcommand{\etc}{\emph{etc.}\xspace}

\begin{document}
\title{Concurrency Control as a Service}


%
\author{Weixing Zhou}
\orcid{0009-0000-9665-6052}
\affiliation{%
  \institution{Northeastern University}
  \country{China}
}
\email{zhouwx@stumail.neu.edu.cn}

\author{Yanfeng Zhang}
\orcid{0000-0002-9871-0304}
\affiliation{%
  \institution{Northeastern University}
  \country{China}
}
\email{zhangyf@mail.neu.edu.cn}

\author{Xinji Zhou}
\orcid{0000-0002-1825-0097}
\affiliation{%
  \institution{Northeastern University}
  \country{China}
}
\email{zhouxj@stumail.neu.edu.cn}

\author{Zhiyou Wang}
\orcid{0000-0002-1825-0097}
\affiliation{%
  \institution{Northeastern University}
  \country{China}
}
\email{wangzy@stumail.neu.edu.cn}

\author{Zeshun Peng}
\orcid{0000-0002-1825-0097}
\affiliation{%
  \institution{Northeastern University}
  \country{China}
}
\email{pengzs@stumail.neu.edu.cn}

\author{Yang Ren}
\orcid{0009-0002-8679-8238}
\affiliation{%
  \institution{Huawei Tec. Co., Ltd}
  \country{China}
}
\email{renyang1@huawei.com}

\author{Sihao Li}
\orcid{0009-0007-4850-4258}
\affiliation{%
  \institution{Huawei Tec. Co., Ltd}
  \country{China}
}
\email{sean.lisihao@huawei.com}

\author{Huanchen Zhang}
\orcid{0000-0002-1825-0097}
\affiliation{%
  \institution{Tsinghua University}
  \country{China}
}
\email{huanchen@tsinghua.edu.cn}

\author{Guoliang Li}
\orcid{0000-0002-1398-0621}
\affiliation{%
  \institution{Tsinghua University}
  \country{China}
}
\email{liguoliang@tsinghua.edu.cn}

\author{Ge Yu}
\orcid{0000-0002-3171-8889}
\affiliation{%
  \institution{Northeastern University}
  \country{China}
}
\email{yuge@mail.neu.edu.cn}

\begin{abstract}
Existing disaggregated databases separate execution and storage layers, enabling independent and elastic scaling of resources. In most cases, this design makes transaction concurrency control (CC) a critical bottleneck, which demands significant computing resources for concurrent conflict management and struggles to scale due to the coordination overhead for concurrent conflict resolution. Coupling CC with execution or storage limits performance and elasticity, as CC’s resource needs do not align with the free scaling of the transaction execution layer or the storage-bound data layer.

This paper proposes Concurrency Control as a Service ({\oursys}), which decouples CC from databases, building an execution-CC-storage three-layer decoupled database, allowing independent scaling and upgrades for improved elasticity, resource utilization, and development agility.
However, adding a new layer increases latency due to the shift in communication from hardware to network. To address this, we propose a Sharded Multi-Write OCC (SM-OCC) algorithm with an asynchronous log push-down mechanism to minimize network communications overhead and transaction latency.
Additionally, we implement a multi-write architecture with a deterministic conflict resolution method to reduce coordination overhead in the CC layer, thereby improving scalability. {\oursys} is designed to be connected by a variety of execution and storage engines. Existing disaggregated databases can be revolutionized with {\oursys} to achieve high elasticity, scalability, and high performance. 
Results show that {\oursys} achieves 1.02-3.11$\times$ higher throughput and 1.11-2.75$\times$ lower latency than SoTA disaggregated databases.
\end{abstract}

\maketitle

\pagestyle{\vldbpagestyle}
\begingroup\small\noindent\raggedright\textbf{PVLDB Reference Format:}\\
\vldbauthors. \vldbtitle. PVLDB, \vldbvolume(\vldbissue): \vldbpages, \vldbyear.\\
\href{https://doi.org/\vldbdoi}{doi:\vldbdoi}
\endgroup
\begingroup
\renewcommand\thefootnote{}\footnote{\noindent
This work is licensed under the Creative Commons BY-NC-ND 4.0 International License. Visit \url{https://creativecommons.org/licenses/by-nc-nd/4.0/} to view a copy of this license. For any use beyond those covered by this license, obtain permission by emailing \href{mailto:info@vldb.org}{info@vldb.org}. Copyright is held by the owner/author(s). Publication rights licensed to the VLDB Endowment. \\
\raggedright Proceedings of the VLDB Endowment, Vol. \vldbvolume, No. \vldbissue\ %
ISSN 2150-8097. \\
\href{https://doi.org/\vldbdoi}{doi:\vldbdoi} 
}\addtocounter{footnote}{-1}\endgroup

\ifdefempty{\vldbavailabilityurl}{}{
\vspace{.3cm}
\begingroup\small\noindent\raggedright\textbf{PVLDB Artifact Availability:}\\
The source code, data, and/or other artifacts have been made available at \url{\vldbavailabilityurl}.
\endgroup
}

\section{Introduction}
\label{section:1}

Database systems are evolving to a compute-storage disaggregated architecture \cite{loesing2015designtell, das2013elastras, huang2020tidb, zhou2021foundationdb, redshift, thusoo2009hive, yu2020pushdowndb, yang2021flexpushdowndb}, such as Amazon Aurora \cite{abadi2003aurora}, Socrates \cite{antonopoulos2019socrates}, PolarDB \cite{cao2021polardb}, and
AlloyDB \cite{alloydb}.
These databases typically decouple the system into an execution layer, which requires substantial computational resources, and a storage layer, which necessitates significant storage capacity.
Compared to traditional databases where execution and storage are bundled together, these two-layer databases allow compute and storage resources to be scaled independently, thereby providing greater elasticity in the cloud environment, which are also called cloud-native databases.
A set of works \cite{guo2022Cornus, yu2020pushdowndb, yang2021flexpushdowndb, zhang2021TowardsSharedMemory} are proposed to improve these cloud-native databases from various aspects.
As more and more enterprises move their applications to the cloud, these disaggregated databases are gaining wide popularity.

\begin{figure}[t]
    \centerline{\includegraphics[width=0.38\textwidth]{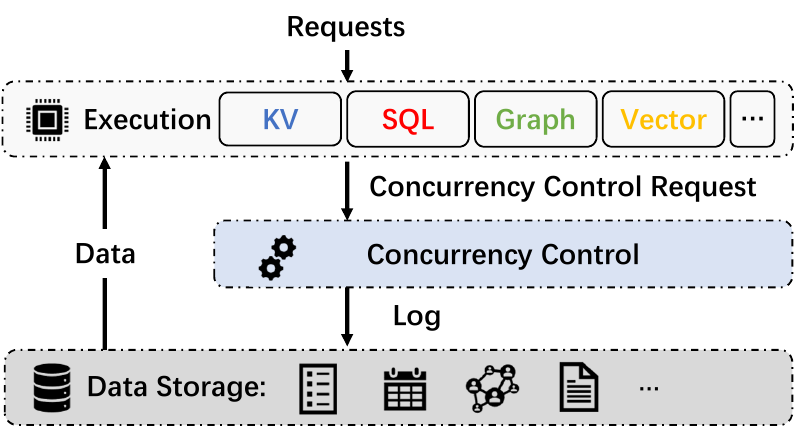}}
    \vspace{-0.13in}
    \caption{An execution-CC-storage three-layer decoupled database architecture.}
    \Description[]{}
    \label{fig:DB}
\end{figure}

The spirit of cloud-native architecture is decoupling. A system should be decoupled into independent function modules, each with specific resource requirements. Cloud provides the elasticity of different decoupled resources (\eg computation, memory, and storage),
allowing the growing or shrinking of resource capacity to adjust to changing demands. Each decoupled function module can be scaled independently to meet varying demands, fully utilizing the decoupled resources. This approach enables the system to achieve elasticity.
Furthermore, these function modules can be designed as independent services \cite{cao2018polarfs, zhu2018solar, googlecloudservice, awss3, alshuqayran2016systematic, cerny2018contextual}, so that each function module can be reused by various applications and can be upgraded independently, thereby bringing more agility. 

Concurrency Control (CC) is a key function module in databases to ensure that concurrent data access operations proposed by different users do not break data consistency. CC deals with concurrent conflicts and guarantees transaction ACID properties, and is evolving towards distributed with the evolution towards highly scalable cloud-native architectures. Adding nodes can enhance distributed transaction CC performance. However, this requires coordination among participates to resolve concurrent conflicts and ensure the ACID properties. System performance drops if the coordination overhead outweighs the benefits of the increased computing resources when adding too much nodes.  
CC has limited scalability, but it still needs high computational resources to resolve concurrent conflicts. The resource requirements of CC are neither consistent with SQL execution nor with data storage. 
Yet, most existing cloud-native databases simply couple CC either with the execution layer \cite{abadi2003aurora, antonopoulos2019socrates, cao2021polardb, loesing2015designtell, das2013elastras} or the storage layer \cite{zhu2018solar, huang2020tidb, taft2020cockroachdb}, which limits the performance and elasticity of these systems (more details are discussed in Section \ref{sec:case}).

Furthermore, the core of CC is resolving read-write conflicts on data items without caring about data types. Such a general applicability is often overlooked by existing databases, which are typically designed for specific engines and data models.

Considering the principle of cloud-native design (\ie decoupling functionality), it is desirable to decouple CC from the database system to maximize scalability and elasticity. By making CC an independent service, it can be connected to multiple engines with different data models (\eg relational, KV, Graph). This approach allows the CC service to be reused more easily and independently upgraded and evolved, promoting development agility.

This paper presents \textbf{\emph{Concurrency Control as a Service} (\oursys)}, a concept aimed at decoupling CC into a separate service and building an execution-CC-storage three-layer database (as shown in Figure \ref{fig:DB}). 
The system can dynamically adjust the resources of each layer based on the workloads (\eg computation, transaction processing, and data storage), thereby improving elasticity and resource utilization. Machines designed for specific scenarios (\eg compute-intensive, parallelism, storage-oriented) achieve better resource utilization by being deployed in different layers.
In this architecture, execution engines independently execute transaction requests, read data from storage, and send resolving requests to \oursys for concurrency control. Once conflicts are resolved, commit or abort notifications are sent back to the execution layer, and committed transaction logs are pushed to the storage layer for data updates. 

Software-level disaggregation results in a significant performance reduction because processing over the network is slower than on a local machine \cite{openAurora2024pang}. Adding a new layer increases latency, as communication between execution and CC is switched from hardware to network.
To tackle this, we propose a Sharded Multi-Write OCC (SM-OCC) algorithm with asynchronous log push-down mechanism. Employing an optimistic execution strategy can reduce the number of network communications between the execution and CC layer, as execution engines no longer need to send locking requests. Asynchronous logging further decreases latency by allowing transactions to commit once logs are persisted in the CC layer, instead of waiting for data updates in the storage layer.  Additionally, the CC layer encounters limited scalability when resolving transaction concurrent conflicts, which necessitates coordinated communication among multiple nodes. To address this, we aim to use a deterministic decision-making method to minimize coordination overhead between nodes to enhance the scalability.

As CC focuses on resolving read and write conflicts on data items, the influence of data models can be mitigated by logically abstracting data access. We design a set of interfaces, which only expose data operation types to the CC layer, so that \oursys can be connected by multiple different engines with various data models.

In summary:
\begin{itemize}[leftmargin=*]
    \item We propose Concurrency Control as a Service (\oursys), a novel execution-CC-storage three-layer decoupled database architecture. CC is decoupled from the database and works as a service, enhancing system scalability, elasticity, and agility. (Section \ref{section:3}). 
    \item We propose a sharded multi-write optimistic concurrency control algorithm (SM-OCC) with asynchronous logging to enhance the scalability of CC and overall system performance (Section \ref{section:4}).
    \item We make several case studies on connecting existing execution/storage engines to \oursys to demonstrate the benefits of the three-layer architecture (Section \ref{section:5}).
\end{itemize}

\section{The case for CCaaS}
\label{sec:case}
\label{section:case}

\begin{figure}
    \includegraphics[width=0.43\textwidth]{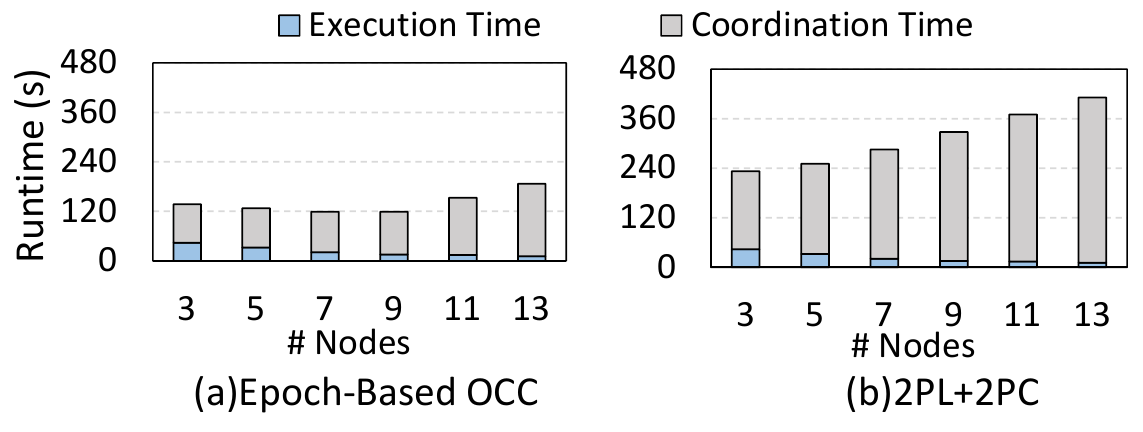}\label{fig:introduction:a}
    \vspace{-0.15in}
    \caption{Breakdown of total processing time for distributed transactions with a changing number of nodes.}
    \Description[]{}
    \label{fig:motivation exp}
\end{figure}

\subsection{Resource Requirements of CC}

To study the resource requirements of distributed CC, we evaluate the runtime of an epoch-based optimistic concurrency control algorithm (epoch-based OCC) and a distributed two phase locking + two phase commit (2PL+2PC) algorithm under a distributed environment. The atomicity and distributed consistency are validated in an epoch-based manner in epoch-based OCC and ensured by 2PC in 2PL+2PC. We generate 10 million distributed transactions using the YCSB-A benchmark, with each transaction containing 10 random operations (5 reads + 5 writes). We process these distributed transactions under different environment setups with different numbers of nodes (3 to 13 nodes), where the data is evenly and randomly distributed among nodes in each run. For more details, please refer to the description of Competitors and the distributed environment setup in Section \ref{Paragraph:Competitors}. We record transaction execution time and inter-node coordination time (\ie waiting for messages from other nodes).

Figure \ref{fig:motivation exp} shows the total time for execution and coordination. No matter epoch-based OCC or 2PL+2PC, most time is spent on inter-node coordination. The total time for transaction execution is steadily decreasing as the number of nodes increases, since more nodes are involved in performing data operations. However, the total time for inter-node coordination is increasing as the number of nodes increases in 2PL+2PC. This is because more nodes could introduce more inter-node coordination overhead. Even though the total runtime is slightly decreasing as the number of nodes increases from 3 to 9 in epoch-based OCC, adding more nodes (greater than 9) can introduce significant coordination overhead and then outweigh the benefits of more compute resources. According to these results, we observe that CC often has limited scaling capabilities due to coordination overhead, and the resource requirement of CC is not consistent with that of transaction execution, where execution prefers relatively more compute nodes but CC prefers fewer nodes. This motivates us to decouple CC from database architecture and make it as an independent service.

\subsection{Limitations of Existing Decoupled DBs}

Most existing decoupled databases overlook the specific requirements of CC and simply couple CC with transaction execution or data storage, leading to performance and scalability limitations.

\begin{figure}[t]
  \centerline{\includegraphics[width=0.47\textwidth]{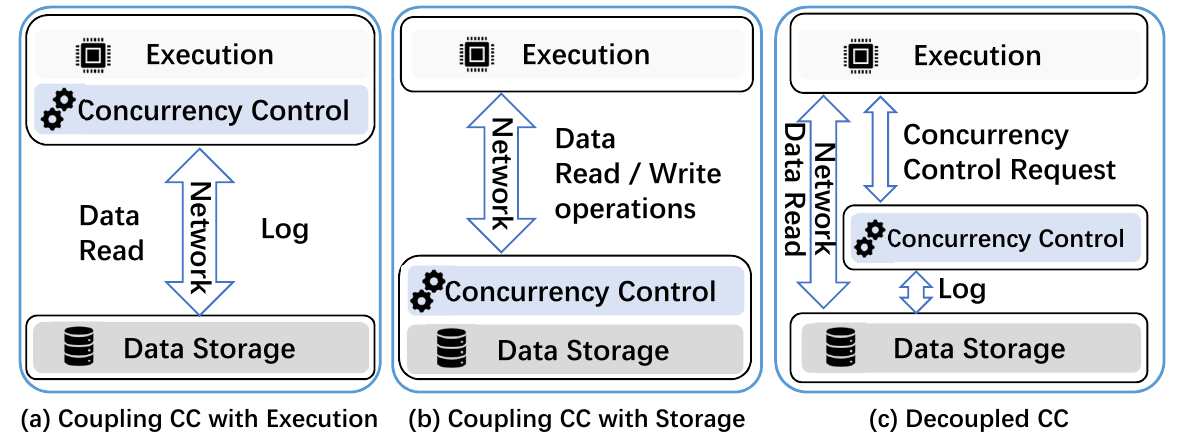}}
  \vspace{-0.13in}
  \caption{Comparison of different decoupled databases.}
  \Description[]{}
  \label{fig:overall-intro}
\end{figure}

Databases like Aurora \cite{abadi2003aurora} and PolarDB \cite{cao2021polardb} \textbf{couple CC with transaction execution}, as illustrated in Figure \ref{fig:overall-intro}a. 
Such a design allows execution nodes to quickly process user requests and manage transaction conflicts. Adding more nodes increases computing resources, distributes the execution load, and prevents single-node bottleneck. However, it also raises coordination overhead by involving more nodes in conflict resolution, scaling too much nodes will hurts the system performance (limited scalability of CC). 
Moreover, coupling CC with execution limits system agility. Execution engines should be tailored to optimize execution plans for various data models or hardware types. For instance, graph databases like Nebula Graph \cite{wu2022nebula} and Neo4j \cite{neo4j} support graph queries such as sub-graph matching and shortest path. Vector databases like Milvus \cite{milvus} are built for similarity search, while GPU databases like GDB \cite{he2009relational}, MapD \cite{mostak2013overview}, and GPUDB \cite{yuan2013yin} are optimized for parallel processing \cite{shanbhag2020study}. However, the core of CC algorithms \cite{yu2018sundial, lu2018star, harder1984observations, yu2016tictoc, thomson2012calvin, lu2020aria, zhou2023geogauss} are similar, \ie handling concurrent read-write or write-write conflicts. Coupling CC with the execution layer would incur redevelopment costs for resolving transaction conflicts.

Some other disaggregated databases, such as Solar \cite{zhu2018solar} and TiDB \cite{huang2020tidb}, \textbf{couple CC with data storage}, as depicted in Figure \ref{fig:overall-intro}b. In these systems, all persistent states reside in the storage layer, rendering execution nodes stateless. The execution nodes are only responsible for computation, forwarding data operations to storage. This design enables dynamic adjustment of execution nodes to efficiently handle varying workloads, optimizing resource utilization. 
Conversely, storage nodes are stateful, and expanding them for more capacity involves tasks like data re-sharding and migration, making them less flexible to scale. Furthermore, storage nodes typically have large amounts of storage space but limited computing resources, and storing large data volumes necessitates numerous storage nodes. 
Coupling CC with storage makes CC hardly elastically scalable. Managing CC with limited computation resources provided by storage nodes could hurt performance.

\section{System Architecture}
\label{section:3}

Motivated by the aforementioned analysis, we build a separate CC service for improving system scalability, elasticity, and agility. 

\subsection{CCaaS Interface}
\label{subsection:Interface}

\Paragraph{{Execution Layer to CCaaS}}
Although data processing in execution engines varies significantly by data models, all have two basic operations: read and write. Based on this, we define unified interfaces that only expose the read and write operations to the CC layer and hide the impact of different data models, for different execution engines to interact with. 

\begin{figure}
      \centerline{\includegraphics[width=0.48\textwidth]{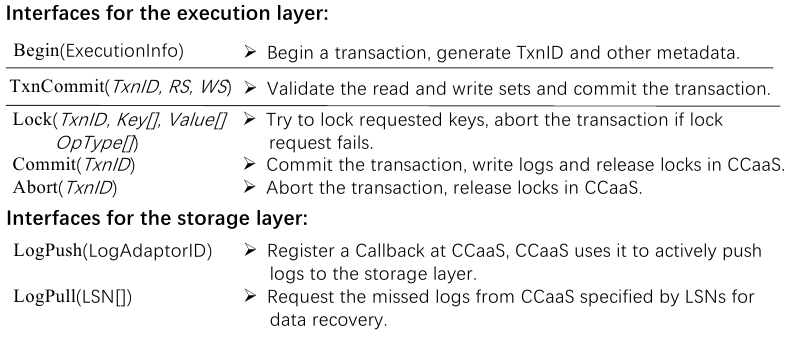}}
      \centering
      \vspace{-0.13in}
      \caption{CCaaS Interfaces.}
      \Description[]{}
      \label{fig:interface}
\end{figure}

For OCC, after transaction execution, an execution engine generates a set of records, including read operations (\ie readset) and write operations (\ie writeset). During CC conflict resolution, the readset and writeset are used to detect read-write conflict and write-write conflict. Therefore, we define a unified interface, \textbf{\texttt{TxnCommit()}}, where execution engines standardize transaction execution results using a defined data structure (Unified Transaction Read Set and Write Set, as shown in Figure \ref{fig:overall_workflow}) and send them to \oursys for conflict detection. On the other hand, for PCC, a set of interfaces (\textbf{\texttt{Lock()}}, \textbf{\texttt{Commit()}, and \textbf{\texttt{Abort()}}}, as shown in Figure \ref{fig:interface}) is provided for the execution layer to interact with \oursys, ensuring correct transaction processing with locks. 

Notably, decoupling CC from existing compute-storage disaggregated databases does not change their ability to support existing data operations. However, it encounters challenges depending on the type of CC algorithms used in \oursys. OCC is an ideal choice for decoupling CC, as it requires only slight modifications to the transaction commit process, which sends the read and write sets of transactions to CCaaS for conflict resolution. Decoupling CC with pessimistic mode (PCC) poses greater challenges and incurs higher costs. Lock requests need to be sent to CCaaS for conflict detection before accessing the data, introducing multiple network round-trips in PCC mode. In cases where the data to be accessed is unknown beforehand, such as accessing data via secondary indexes, additional operations (e.g., an initial optimistic execution to identify the data) are required. In this paper, we mainly focus on supporting OCC as a first step toward achieving a decoupled CC service.

\Paragraph{{CCaaS to Storage Layer}}
Storage engines (\eg columnar \cite{abadi2008column}, row-column hybrid \cite{he2011rcfile}, graph-native \cite{deutsch2019tigergraph}, and vector \cite{wang2021milvus}) differ significantly in their data structures and update interfaces, pre-developing interfaces in \oursys to update data for all storage engines is impractical. 
Thus, we choose to \textbf{transform the write sets of transactions into logs} and send the logs to the storage layer. Meanwhile, a \textbf{LogAdaptor} is required for a storage engine to provide a specification on how to transform the logs to the corresponding data with a specific schema (Section \ref{subsubsection:Logging}). We define a Callback \textbf{\texttt{LogPush()}} that is registered at \oursys, so that \oursys will use it to push logs to the storage layer actively, and the LogAdaptor receives these logs and converts them to data. Furthermore, the storage layer relies on the \textbf{\texttt{LogPull()}} interface to verify the completeness of the logs received according to the log sequence number (LSN) and requests missed logs from \oursys.

\vspace{-0.05in}
\subsection{System Workflow}
\begin{figure}
      \centerline{\includegraphics[width=0.4\textwidth]{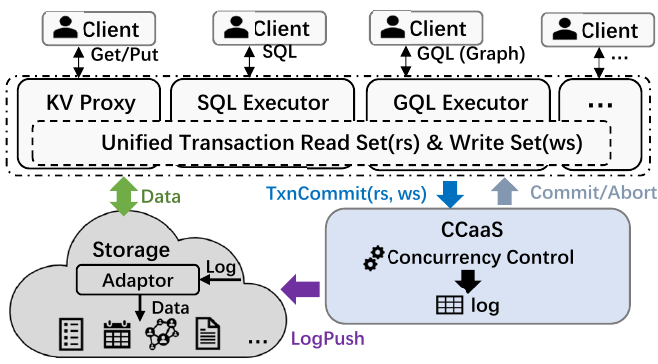}}
      \centering
      \vspace{-0.15in}
      \caption{The overall architecture of CCaaS (OCC-based).}
      \Description[]{}
      \label{fig:overall_workflow}
\end{figure}

By connecting execution and storage engines to {\oursys}, we construct an execution-CC-storage three-layer decoupled database. 
The execution layer, which may consist of multiple engines with different data models (\eg KV, Relational, Graph), receives user requests, executes computation, reads from the storage, and sends unified CC requests to \oursys for conflict resolution. It is noticeable that for KV engines that lack transaction support, such as LevelDB \cite{leveldb} and HBase \cite{hbase}, only get/put operations are supported on a single item. An additional transaction proxy (\eg the KV proxy as shown in Figure \ref{fig:overall_workflow}) is required to provide transaction semantics (\eg \texttt{BeginTransaction}, \texttt{Commit}, \texttt{Abort}) for users to interact with. 

When a transaction request arrives, the execution layer calls the \texttt{Begin()} interface to initialize a transaction. In OCC, the execution layer optimistically executes the transaction. For example, SQL engine parses SQL statements (\eg SELECT) and retrieves metadata (\eg table schema, data distribution) from storage to generate a physical execution plan. 
Based on the plan, the engines read data from the storage layer without locking (stale reads may occur) and cache reads and writes locally.
When transaction execution is finished, the execution layer sends the read and write sets to \oursys for conflict detection. In PCC, \oursys provides a distributed lock manager for lock authorization and conflict detection. Unlike OCC, PCC requires several interactions between the execution engines and \oursys. When the execution layer invokes the \texttt{Commit()} interface in PCC or after resolving conflicts in OCC, \oursys returns Commit/Abort to the execution layer and pushes the updates in the format of logs to the storage layer through the \texttt{LogPush} interface.

Since OCC requires much fewer interactions between the execution layer and the CC layer, OCC is more suitable as the CC protocol in \oursys.
To enhance CC scalability and performance, we propose a \textbf{Sharded Multi-Write OCC (SM-OCC)} algorithm as the default CC mechanism in \oursys in Section \ref{section:4}.

\section{CCaaS Design}
\label{section:4}
\label{section:CCaasLayerDesign}

\subsection{CCaaS Overview}
\label{subsection:CCaaS architecture}

\Paragraph{Requirements}
Since the CC layer operates as an independent service, \oursys should satisfy a set of specific requirements:

First, \oursys must be highly available. Classic master-follower architecture provides high availability but comes with several significant drawbacks. 1) All the conflicts must be resolved in the master node; a single-node bottleneck occurs when concurrent CC requests increases, causing heavy resource contention. 2) The master-follower architecture experiences temporary unavailable when the master node goes down unexpectedly. The system necessitates a pause for complete log replaying in the followers. To address this challenge, we adopt a \textbf{Multi-Master} architecture \cite{Depoutovitch2023Taurus, Yang2024PolarDB, AuroraMM}, where each master has the capacity to resolve transaction conflicts. 
CC requests are distributed across multiple masters, preventing single-point performance bottlenecks. If a master fails, other masters continue providing CC services without interruption.

\begin{figure}
        \centerline{\includegraphics[width=0.45\textwidth]{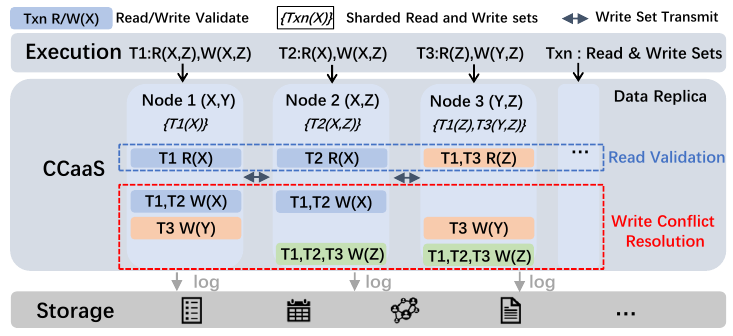}}
        \centering
        \vspace{-0.15in}
        \caption{The architecture of CCaaS with SM-OCC.}
        \Description[]{}
        \label{fig:CCaaS_architecture}
\end{figure}

Second, while the use of a multi-master architecture can fully utilize each replica and provide high availability, it incurs a write amplification problem. The same set of writes needs to be processed in every master to maintain replica consistency. 
To mitigate this, \textbf{Sharding} strategies \cite{Clay2016Serafini, das2013elastras, E-store2014Taft, Accordion2014Serafini, Schism2010Curino} can be used to enhance the scalability, where each node handles only a subset of shards, reducing the write amplification impact.

Third, the transaction execution should be \textbf{Optimistic}. Prior studies have proposed various concurrency control algorithms, including pessimistic \cite{yu2018sundial, lu2018star, harder1984observations}, optimistic \cite{yu2016tictoc, tu2013speedy, lu2021epoch, zhou2023geogauss}, and deterministic \cite{thomson2012calvin, lu2020aria, ren2019slog, qadah2020qstore, hildred2023caerus} methods, each with its own pros and cons. Using a PCC algorithm in the CC layer involves sending numerous lock requests from the execution layer, leading to increased network traffic and transaction latency. Deterministic algorithms provide strong consistency guarantees among multiple replicas through \textit{deterministic execution}, which reduces coordination overhead. However, they are not universally applicable, as they struggle with interactive transactions and often require pre-processing.
To leverage the strengths of both optimistic and deterministic approaches, we propose a combined method. Transactions are \textbf{\textit{optimistically executed}} in the execution layer. Once a transaction is submitted to \oursys, the multi-master nodes in the CC layer \textit{\textbf{deterministically compare}} the read and write sets according to predefined rules.
It is worth noting that deterministic comparisons occur only during the validation phase. Users can still interact with the database during transaction execution. Support for multi-round and interactive transactions remains unaffected.

Fourth, \textbf{Epoch-Based Committing} suits the sharded architecture better. Typically, sharded systems use protocols like two-phase commit (2PC) for transaction atomicity, which requires multiple round-trip acknowledgments, leading to significant network overhead. The Epoch-Based Commit protocol \cite{lu2021epoch} groups transactions into epochs, using the entire epoch as the coordination unit to minimize communication overhead. The original transaction-granularity synchronization is transformed into epoch-granularity, effectively reducing coordination overhead.

Given these, we propose a \textbf{Sharded Multi-Write OCC (SM-OCC)} algorithm.
Notably, the CC algorithm (\eg SM-OCC) is changeable. Lock-based algorithms like 2PL can also be used in \oursys to reduce the abort rate caused by using OCC. If the storage layer supports multi-version data, MVCC can also be integrated into \oursys \cite{loesing2015designtell, ferro2014omid}. Algorithms \cite{Chen2017Fast, fast2015wei, Chen2016Fast, Wei2018Deconstructing, fast2023du, fasst2016kalia} that leverage RDMA to reduce network overhead can also be applied to \oursys to enhance performance.
Users can choose an appropriate CC algorithm that suits their workloads but must consider its impact on system availability, scalability, and performance.

\begin{figure}
    \centerline{\includegraphics[width=0.5\textwidth]{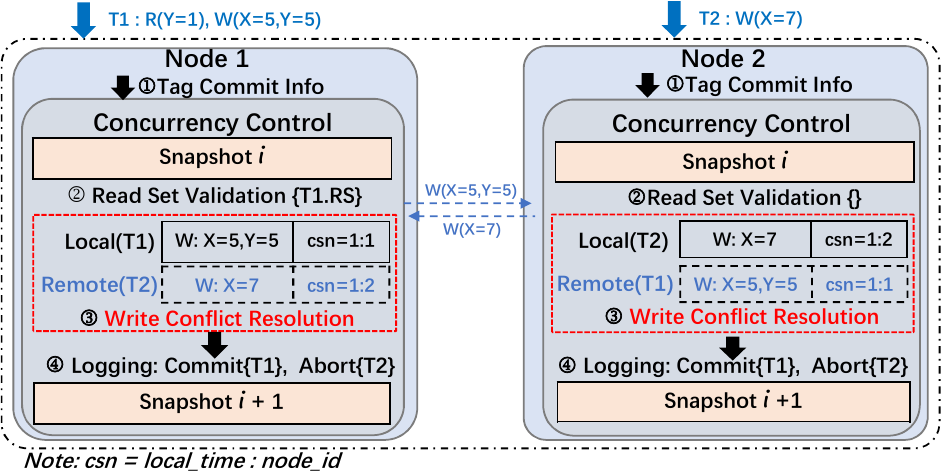}}
    \centering
    \vspace{-0.1in}
    \caption{Conflict resolution within a data shard.}
    \Description[]{}
    \label{fig:Epoch-Based-Multi-Master-transactionprocessing}
\end{figure}

\Paragraph{{CCaaS Architecture with SM-OCC}}
In \oursys, each node maintains several shards of committed transaction metadata. For example, as shown in Figure \ref{fig:CCaaS_architecture}, node 1 manages Shards $X$ and $Y$, node 2 manages Shards $X$ and $Z$. Identical replicas are deployed across nodes, using the Raft \cite{raft} consensus protocol to synchronize updates and ensure consistency. Each node acts as a master, allowing nodes to manage transaction conflicts independently (\eg both node 1 and node 2 can resolve conflicts on Shard $X$). They receive read and write sets (referred to as transactions) and partition them based on a sharding strategy (\eg range-based or hash-based), routing subtransactions to corresponding nodes for conflict resolution. Since each node has the capacity for conflict resolution, subtransactions can be routed to any one of the nodes (masters). 
For instance, node 1 shards transaction $T1$ into $T1(X)$ and $T1(Z)$, sending $T1(Z)$ to Node3. Alternatively, $T1(Z)$ can also be send to node 2 since it holds a replica of $Z$.

\oursys provides a certain degree of scalability and elasticity. When scaling out for more computational resources, \oursys assigns shard replicas to the new node. After synchronizing metadata with peers, the new node begins resolving transaction conflicts. When scaling in, nodes transfer the replicas to other nodes for replacement. If the access rate of a shard increases, resulting in higher resource contention on some nodes, \oursys can share the overhead of read-set validation by increasing the number of replicas, or re-partition the shard to distribute the overhead of write-set resolution. CCaaS only maintains the meta-information of committed transactions, and when re-sharding is performed, only the meta-information managed by the CC nodes is redistributed. \oursys does not experience significant network bandwidth usage.

\subsection{Sharded Multi-Write OCC}
\label{subsection:Transaction Management}

\Paragraph{{Multi-Write OCC}} 
For convenience, we first introduce conflict resolution within a shard and then present the difference with sharding.
The main workflow are shown in Figure \ref{fig:Epoch-Based-Multi-Master-transactionprocessing}. First of all, \oursys divides physical time into epochs (\eg 10 ms per epoch) and assigns incremental unique numbers to these epochs. Each node collects read and write sets (transactions) and packs them at epoch granularity based on the reception time. Upon receiving a transaction, \oursys node tags it with commit-info, which includes the commit epoch number (CEN, indicating the epoch to which the transaction belongs) and the commit sequence number (CSN, identifying the transaction), for subsequent processing.
 
Nodes synchronize transactions with each other at epoch granularity, and operate in an epoch manner: after transactions of the $i$ epoch have been executed, snapshot $i$ is generated and nodes start the conflict resolution for transactions of epoch $i + 1$ (\eg $T1, T2$ in node 1 and node 2).

There are two types of conflicts between transactions: read-write and write-write conflicts. The core procedure of SM-OCC is divided into \textbf{Read Set Validation} and \textbf{Write Set Resolution} phases. 

\begin{figure}
    \hspace{-0.3in}\centerline{\includegraphics[width=0.44\textwidth]{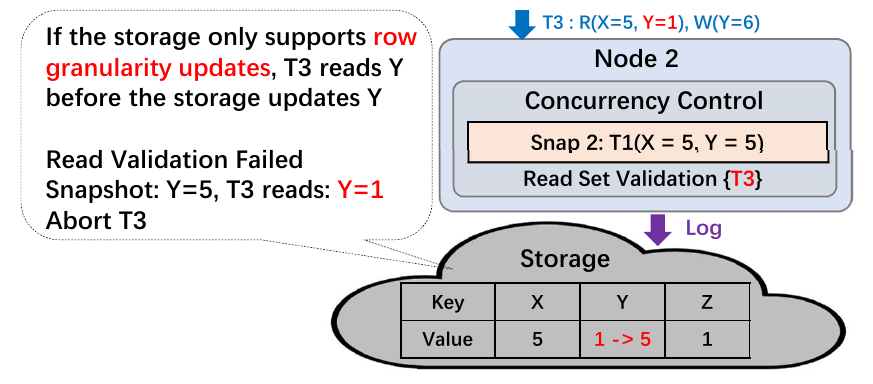}}
        \centering
        \vspace{-0.15in}
        \caption{Read validation when connecting to the storage engines only supports row granularity updating.}
        \Description[]{}
        \label{fig:ReadSetValidation}
\end{figure}

\Paragraph{{Read Set Validation}}
\label{Paragraph:Read Set Validation}
Each node first independently validates the read sets of locally received transactions based on snapshot $i$. \oursys adopts Snapshot Isolation (SI) by default to validate the reads.
When \oursys connects to storage engines with only row-level updates, an additional read-error scenario may occur: a transaction may read data partially modified by another transaction, beyond the conflicts seen with traditional transaction-level updates. Figure \ref{fig:ReadSetValidation} shows this issue. This read error occurs when transaction $T1$ updates $Y$, which has already been read by $T3$. $T1$ updates $X$ and $Y$ to 5 and commits. $T3$ reads $X$ as 5 and $Y$ as 1. Despite no conflicts in the same epoch, $T3$ must be aborted to maintain atomicity (\ie it should read all or none of $T1$'s updates). 
To ensure atomicity, transactions like $T3$ that are aborted by checking snapshots.

\begin{algorithm}[t]
\begin{spacing}{1}
\SetAlgoNoLine
\small
    \caption{Write Set Conflict Resolution}
    \Description[]{}
    \label{alg:Write Resolution}
    \KwIn{a transaction $txn.\{CSN, WS\}$.}
    \KwOut{Commit or Abort}
    \SetKw{Continue}{continue}

    \SetKwFunction{FWriteSetResolution}{WriteSetResolution}
    \SetKwFunction{FCompare}{Compare}

    \SetKwProg{Fn}{Function}{:}{\KwRet}
    \Fn{\FWriteSetResolution{Transaction $txn$}}{
        $result = Commit$\;
        \ForEach{$r$ in $txn.WS$}{
            $row$ = GlobalWriteVersionMap.Find($r$.key)\;
            \uIf{$row$ is not $Null$ \textbf{and} $r.type$ is $Insert$}{
                    $result = Abort$; \textcolor{gray}{//row already exists.}\\
            }
            \uElseIf{$row$ is $Null$ \textbf{and} $r.type$ is not $Insert$}{
                    $result = Abort$; \textcolor{gray}{ //deleted in previous epoch.}\\
            }
            \uElse{
                $result$ = Compare($r$)\;
            }
        }
        \uIf{$result$ is $Abort$} {
            EpochAbortSet.Insert($txn.CSN$)\;
        }
        \uElse {
            \ForEach{$row$ in $txn.WS$}{
                GlobalWriteVersionMap.Set($row$)\;
            }
        }
        \KwRet $result$\;
    }

    \SetKwProg{Fn}{Function}{:}{\KwRet}
    \Fn{\FCompare{Record $r$}}{
        $row$ = EpochWriteVersionMap.Find($r$.key)\;
        \uIf{$row$ is $Null$}{
           \textcolor{gray}{ //row has not been updated in current epoch.}\\
            $row.CSN = txn.CSN$\;
            EpochWriteVersionMap.Set($row$)\;
        }
        \uElseIf{($row.CSN > txn.CSN$)} {
                \textcolor{gray}{ //mark the transaction with CSN $row.CSN$ as abort.}\\
                EpochAbortSet.Insert($row.CSN$)\;
                $row.CSN = txn.CSN$\;
                EpochWriteVersionMap.Set($row$)\;
            }
            \uElse{
                \KwRet $Abort$\;
            }
        \KwRet $Commit$\;
    }
\end{spacing}
\end{algorithm}

\Paragraph{{Write Set Conflict Resolution}}
\label{Paragraph:Write Set Conflict Resolution}
For transactions passing the read validation phase, their write sets are sent to remote nodes (other masters) to ensure data consistency between replicas. Once all local write sets of epoch $i+1$ are sent and all peers' write sets are received (epoch synchronization), the write resolution phase begins.

Algorithm \ref{alg:Write Resolution} is designed as \textbf{deterministic} to resolve write conflicts, allowing each node to resolve conflicts independently without coordination, thus reducing communication overhead. Each node uses a \textit{GlobalWriteVersionMap} (snapshot) to record information of committed transactions. For each epoch, each node uses an \textit{EpochWriteVersionMap} to track transaction write intents within the current epoch and maintains an \textit{EpochAbortSet} to store the CSNs of transactions that should be aborted.

During the resolution phase, records (\eg rows) in the write set are traversed to detect conflicts. 
For each write operation, it is essential to first check its validity: operations that attempt to update or delete a non-existing row, or insert a row that already exists, are not allowed as they conflict with committed transactions. To verify this, the \textit{GlobalWriteVersionMap} is first searched to determine the row's current state, ensuring the write operation can proceed correctly at the storage layer (lines 5-8).

If no conflicts with committed transactions are found, the \texttt{Compare} function (lines 17-30) is used to check for write conflicts within the current epoch. In the function, the \textit{EpochWriteVersionMap} is used to detect if another transaction is attempting to update the same row
: 1) $row == NULL$ means that no other transaction has tried to update the row, so the current transaction can proceed. The updating intent with the transaction's CSN is inserted into the map(lines 16-19).
2) Otherwise, a write-write conflict occurs, and a partial order `$\prec$' between CSNs is used to determine which transaction wins (lines 20-26). 
The CSN of a transaction is composed of local time + node id. To avoid single-point bottlenecks caused by using a central sequencer, the CSN is assigned using local timestamp with node id.
\begin{definition}
$T1$ `$\prec$' $T2$ if `$T1.local\_time < T2.local\_time $' or `$T1.local\_time = T2.local\_time$ $\&$ $T1.node\_id < T2.node\_id$'.
\end{definition}

Notably, no two transactions with the same CSN exist. The case of $row.CSN == txn.CSN$ will never occur by using the local clock with node ID to assign CSN.
Moreover, the comparison rules are changeable. The rules mentioned above lead to the fact that transactions arriving on nodes with smaller local times (clock skew) have a higher probability of winning, which may lead to inequities. For example, the comparison rules can be changed to commit transactions based on polling, making the comparison a bit fairer. 

After completing the conflict resolution of all transactions, each node updates the \textit{GlobalWriteVersionMap}, and writes logs. Then, a new snapshot based on the current epoch is generated. 

To reduce data replication overhead, each node can performs write-set conflict resolution on local received transactions (local) first and only send the write-sets of the winning transactions to peer nodes. In this way, part of the write conflict resolution load can be distributed among nodes (masters), improving \oursys scalability. Since the comparison rule is deterministic, transactions that lose in local conflict detection must be aborted even if their write-sets are sent to remote nodes. Performing local resolution and only sending wining write-sets does not affect the correctness of \oursys.

\Paragraph{{Sharding}}
\begin{figure}
        \centerline{\includegraphics[width=0.47\textwidth]{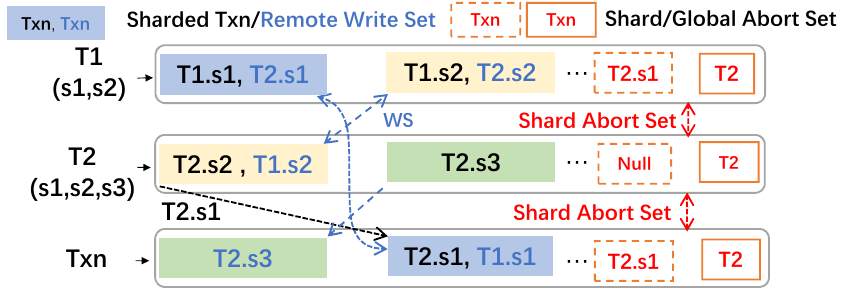}}
        \centering
        \vspace{-0.13in}
        \caption{Transaction processing in sharded multi-write OCC.}
        \Description[]{}
        \label{fig:transactionprocessing-sharding}
\end{figure}
Figure \ref{fig:transactionprocessing-sharding} shows the sharded architecture with three nodes, forming a three-shard, two-replica setup. In this setup, each \oursys node manages a portion of the read and write conflicts (\eg node 1 and node 3 are both responsible for shard 1, blue shard). Transactions are sharded into subtransactions and rerouted to the corresponding nodes. For instance, node 1 shards $T1$ into $T1.s1$ and $T1.s2$. Since node 1 hosts replicas of shard 1 and shard 2, it validates both sub-transactions locally. In contrast, node 2 sends $T2.s1$ to node 3, as node 2 does not maintain a replica of shard 1.

Once conflict resolution is completed, each node generates an \textbf{EpochAbortSet} (Shard). Since each node only manages part of the sub-transactions, the AbortSet on each node may do not account for all transactions that need to be aborted. 
Therefore, an additional round of EpochAbortSet replication is needed to construct a \textbf{globally consistent abort set} for the epoch to ensure transaction atomicity. Then, each node aborts the relevant transactions based on the globally abort set and logs the committed write sets.

\subsection{\textbf{Isolation}}
\label{Paragraph:Isolation}
\oursys uses an epoch-based mechanism and defaults to Snapshot Isolation (SI). For storage engines with multi-version read support, read set validation can be skipped to speed up conflict resolution. 
At the start of a transaction, a start timestamp is recorded for validation. Upon commit, the transaction’s read and write sets, along with the timestamp, are sent to \oursys. \oursys uses this timestamp to determine the appropriate snapshot for validation. Unlike traditional methods of preventing phantom reads, such as using index locks, \oursys detects phantom reads based on the transaction’s execution result. If the read version does not match the snapshot, the transaction is aborted.

\oursys also supports Read Committed (RC) and Repeatable Read (RR) levels when the storage engine supports transaction-level updates. 
However, achieving Serializable (SER) isolation would require global read-write dependency tracking across nodes and epochs, introducing high network and computation overhead, which limits scalability. Considering these performance concerns, CCaaS currently does not support this isolation level. A possible approach is to build a read-write dependency graph and break dependency cycles \cite{serializablesnapshotisolation}.

\subsection{\textbf{Log Pushing}}
\label{subsubsection:Logging}
As discussed in Section \ref{section:3}, LogAdaptor converts logs into data structures compatible with each storage engine. To minimize engine modifications, the adaptor is preferably implemented in \oursys. For example, for HBase \cite{hbase}, we implement the adaptor in \oursys, which converts logs into structures (rowArray, rowOffset, and rowLength), and then invoke HBase's interface \texttt{Put} to update data. For engines without direct data modification support, modifications to the database are required. For instance, openGauss \cite{opengauss}, which only provides SQL interfaces for updates, the log adaptor is implemented within openGauss, invoking internal update mechanisms to apply changes.

During logging, \oursys first writes logs to local disks for persistence before pushing them to the storage layer and returning resolution results to the execution layer. When using SM-OCC, an asynchronous log-pushing mechanism can be used to reduce committing latency, where each node returns the resolution results to the execution layer before finishing pushing the log. This does not affect correctness, as \oursys maintains up-to-date meta-information for conflict resolution. However, in write-intensive workloads, it may increase transaction abort rates due to stale data reads.

\subsection{Fault Recovery}
\label{subsection:Fault Recovery}

Building a execution-CC-storage three layered database requires careful handling of failures at each layer.

\Paragraph{\textbf{Failure in the execution layer}}
Execution nodes optimistically read data and temporarily store updates in memory. If a node fails before committing the transaction to \oursys, the updates are lost, but the other layers remains unaffected. The user can resubmit the transaction to another active node.
If a node fails while waiting for the CC result, the user can connect to another node to check if the data has been updated and confirm if the transaction was committed. Since the transaction is already sent to \oursys, it will be correctly processed without effectiveness.

\Paragraph{\textbf{Failure in the \oursys layer}}
\oursys uses the Raft protocol \cite{ongaro2014searchraft} to ensure fault tolerance. In SM-OCC, each shard master maintains a Raft instance (Shard Raft) for to make a consensus on the status of live masters, which can prevent permanent blocking (waiting for the write sets from a failed master). Additionally, each node maintains a Raft instance (Txn Raft) to replicate its locally received transactions to other nodes for backup. Once Txn Raft replication completes, nodes proceed with transaction sharding and conflict resolution.

When a node fails, \oursys takes different actions depending on whether the node completed the write set exchange for the current epoch. For simplicity, consider a non-sharded example: node 1 and node 2 receive transactions $T1$ and $T2$ respectively, which have a write-write conflict. If node 1 fails before $T1$'s write set is replicated, node 2 produces a GlobalAbortSet without $T1$. During log push, the newly elected Txn Raft leader checks the GlobalAbortSet and pushes $T1$'s log, leading to inconsistencies as both $T1$ and $T2$ are pushed to the storage layer. To prevent this, \oursys re-executes the CC for the current epoch if a failed node hasn’t completed write set replication. The newly elected Txn Raft leader takes over conflict resolution for the failed node and transmits the write sets of the transactions received by the faulty node to other nodes.

In a sharded setup, incomplete EpochShardAbortSets from different masters can result in an incorrect GlobalAbortSet, necessitating epoch re-execution. When a node fails, new Raft leaders are elected. The new Shard Raft leader takes response for the failed node.
If the write set transfer was completed before the failure, other shard masters can process the complete write sets for the current shard and produce a correct conflict resolution result (\ie EpochShardAbortSet). If not, \oursys will re-execute the current epoch with the updated Raft leaders.

\oursys’s multi-master architecture ensures that individual node failures do not affect availability. Execution nodes connected to failed nodes can reroute to healthy ones, ensuring uninterrupted CC service. When the failed node recovers, Raft leadership is restored, and the system returns to normal. In cases where multiple nodes in a Raft instance fail, conflict resolution cannot proceed without achieving Raft consensus. \oursys responds by re-sharding the data, rebuilding Raft instances, and re-executing the epochs. Even if a majority of \oursys nodes fail due to network partitions or power outages, the database system can still provide read-only service by passing \oursys. In such cases, no Raft leader is elected, log push-down is terminated. Upon recovery, nodes first push their remaining local logs. As all states stored in main memory are lost, the nodes then catch up with the latest state by fetching Raft logs before resuming service.

\Paragraph{\textbf{Failure in the storage layer}}
As described in Section \ref{section:3}, the storage layer may include distributed systems like HBase \cite{hbase} or standalone databases like openGauss \cite{opengauss}. Distributed systems already provide fault tolerance, so no additional mechanisms are needed. For standalone databases, multiple instances must be deployed, each with a full replica (see Section \ref{subsection:Making Standalone TP Distributed}). The failure of a majority of storage nodes does not affect the storage layer’s ability to serve data. When a node recovers, it pulls and replays logs from \oursys or peer nodes.

\section{Discussion}
\label{sec:5:discuss}

\Paragraph{{Opportunities of Decoupling}}
\label{Paragraph:Opportunities}
Most database developers tend to decouple database systems based on the disaggregation of storage and computing provided by cloud providers, often prioritizing hardware disaggregation while overlooking resource contention among functional modules. Considering the principles of decoupling, the performance implications of module coupling need to be considered. For example, CC requires efficient handling of concurrent transactions, while logging favors sequential I/O, and record storage demands efficient random access. In serverless architectures, different query operators exhibit varying computational demands, necessitating tailored resource allocation (\eg parallel scanning vs. high-frequency processing).

Another key aspect is functional generality. Decoupled function modules can be designed as generalized services, \eg CCaaS not only resolves conflicts across execution nodes but also supports heterogeneous execution engines, facilitating a heterogeneous execution layer (Section \ref{subsection:Cross-Model Transaction}). Additionally, CCaaS ensures broad compatibility through log adaptors, allowing integration with various storage engines (e.g., columnar, row-based, disk, or memory) based on system requirements.

\begin{figure}[t]
  \centerline{\includegraphics[width=0.47\textwidth]{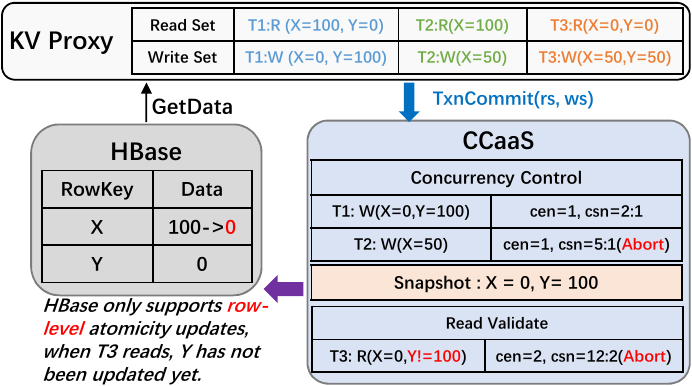}}
  \centering
  \vspace{-0.1in}
  \caption{An example of empowering HBase with TP capability.}
  \Description[]{}
  \label{fig:ap-tp}
\end{figure}

\Paragraph{{Trade-offs}}
\label{Paragraph:Trade-offs} 
Decoupling CC offers several benefits, and we demonstrate several case studies in Section \ref{section:Case Studies} for illustration. 
However, the impact on system performance and maintenance complexity should also be considered. Decoupling CC introduces additional network communication overhead, and CCaaS is more suitable for OCC. Decoupling CC in the case of using PCC algorithms tends to incur higher overhead.
Furthermore, decoupling may lead to increased overhead due to the need to manage the three distinct layers. Each layer will have its own maintenance requirements, which could increase maintenance complexity. Additionally, in highly partitioned workloads, coordination overhead across database nodes is minimal. The scalability of the system can not be limited by coupling CC with other layers. In such situations, decoupling CC for higher scalability is not necessary.

\Paragraph{{Implications to Existing Cloud-Native Databases}}
\label{Paragraph:Implications}
Existing cloud-native databases adopt the \textit{Log-as-the-Database} principle to reduce network I/O. CCaaS and the SM-OCC algorithm can seamlessly integrate with the storage-disaggregated architectures \cite{openAurora2024pang}, introducing several key improvements: CCaaS enables traditional master-follower architectures to evolve into multi-master architectures, improving elasticity and availability. Execution nodes can retain their local CC mechanisms while CCaaS enforces global consistency, allowing for a flexible concurrency control design that adapts to different workload characteristics. Furthermore, SM-OCC employs row-level conflict resolution, improving parallelism compared to page-level approaches \cite{Yang2024PolarDB, Ziegler2022ScaleStore, Depoutovitch2023Taurus}. With CCaaS, existing databases can gain more flexibility in system architecture and performance optimization.

\section{Case Studies}
\label{sec:5}
\label{section:5}
\label{section:Case Studies}

In this section, we demonstrate several advantages of \oursys by presenting some study cases. 

\subsection{Empowering NoSQL DBs with TP Capability}
\label{subsection:Empowering NoSQL DBs with TP Capability}

Most NoSQL databases \cite{leveldb, hbase, wu2022nebula, neo4j} do not provide transaction semantics for better scaling. We offer a solution to add transaction support without modifying the original logic. By connecting existing NoSQL databases to \oursys, these databases can be empowered with TP capability and ACID properties. For example, Figure \ref{fig:ap-tp} shows how we connect HBase, a distributed key-value store, to \oursys to enable TP. First, we implement a KV Proxy to provide transaction semantics, linking it with both \oursys and HBase. The proxy provides \texttt{Start, Get, Put, RollBack} and \texttt{Commit} interfaces for transaction operations and executes user read/write requests, and caches the data locally. When a transaction is committed, the proxy sends the cached read and write sets to \oursys. As described in Section \ref{subsubsection:Logging}, a log adaptor in \oursys uses HBase’s putRow interface to update the data.
We use HBase \cite{hbase} as an example, and \oursys has also been connected to NoSQL DBs like LevelDB \cite{leveldb} and NebulaGraph \cite{wu2022nebula}. 

\subsection{Making Standalone TP Distributed} 
\label{subsection:Making Standalone TP Distributed}

\begin{figure}[t]
    \centerline{\includegraphics[width=0.47\textwidth]{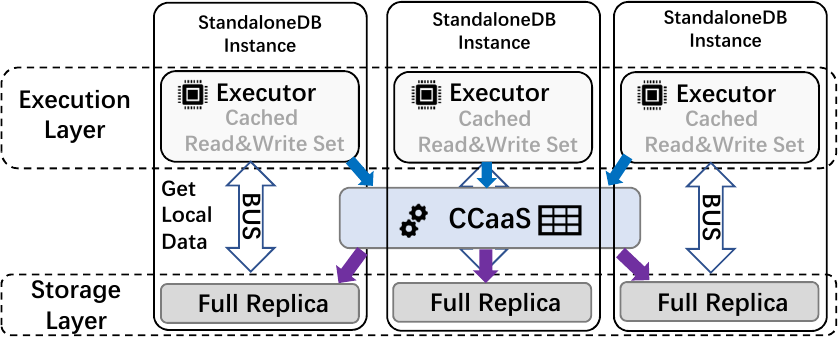}}
    \centering
    \vspace{-0.13in}
    \caption{An illustration of building a multi-master DB.}
    \Description[]{}
    \label{fig:multimaster}
\end{figure}

The performance of standalone databases like openGauss \cite{opengauss}, PostgreSQL \cite{postgre}, and MySQL \cite{mysql} is inherently constrained by the scalability limits of single-node hardware. These systems struggle under high TP workloads and face availability challenges due to their single-node deployment model.
By connecting multiple standalone instances to \oursys, a multi-master distributed TP system can be easily established (Figure \ref{fig:multimaster}). Each instance maintains a full replica of data, independently processes users' requests, and sends CC requests to \oursys to resolve transaction conflicts within and between instances. 
Compared to the master-follower architecture, this design can maximize resource utilization across all instances, distribute the workload, and eliminate single-node bottlenecks. The system also achieves high availability, as other instances can continue serving user requests when some fail.

\subsection{Supporting Cross-Model Transactions}
\label{subsection:Cross-Model Transaction}

Modern business involves multiple data models (\eg Relational, KV, Graph, Vector) stored in various databases, often adopting different storage engines for diverse data needs. 
Businesses may need a transaction with ACID properties to modify data in different databases. 
We assume that a cross-model database is a database that can store, index, and query data across multiple data models. However, ensuring ACID transactions across these databases is challenging due to heterogeneous query languages and the complexity of maintaining consistency across multiple data stores \cite{dey2015scalabletCherryGarciaprotocol, ferro2014omid, zhang2022skeena, kraft2023epoxy}.

\begin{figure}[t]
    \centerline{\includegraphics[width=0.48\textwidth]{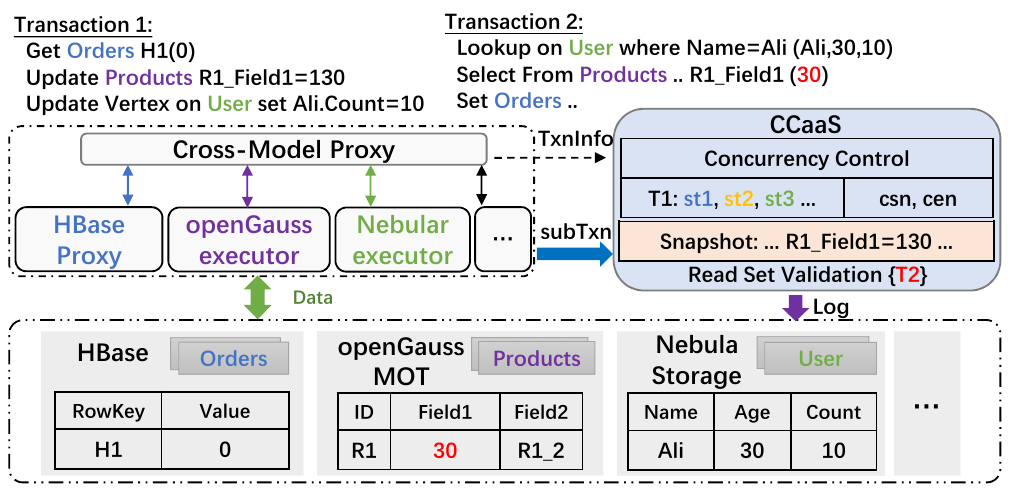}}
    \centering
    \vspace{-0.17in}
    \caption{Cross-model transaction processing with CCaaS.}
    \Description[]{}
    \label{fig:multimodeltransaction}
\end{figure}

\oursys provides a trivial solution by decoupling concurrency control from data models, enabling cross-model transaction support. 
As shown in Figure \ref{fig:multimodeltransaction}, users send requests to a Cross-Model Proxy, which forwards them to relevant execution engines, splitting a cross-model transaction into multiple single-model sub-transactions. 
Once all single-model transactions are executed, the \textsf{Proxy} sends commit commands to each execution engine to submit single-model transactions to \oursys. Meanwhile, it sends transaction information to \oursys, indicating which single-model transactions belong to the same cross-model transaction.
Upon receiving all sub-transactions and metadata, \oursys merges them and applies the same conflict resolution algorithm, ensuring consistency across heterogeneous storage engines.

Since different storage engines cannot directly communicate, updates in a cross-model transaction cannot occur atomically across engines.
This may lead to inconsistent reads in new transactions (e.g., $T2$ in Figure \ref{fig:multimodeltransaction}), violating transaction atomicity. 
\oursys detects such anomalies by validating $T2$’s read set against snapshots and aborts $T2$ if inconsistency is found. After resolving concurrency conflicts, \oursys returns the results to the Proxy and execution nodes, then pushes logs to the respective storage engines. Since cross-model transactions span multiple engines, each log entry is labeled with an identifier specifying its target engine, ensuring correct log propagation and consistency.

\section{Evaluation}
\label{section:6}
\label{section:Evaluation}

This section evaluates the performance of \oursys with SM-OCC. 


\Paragraph{\textbf{Implementation}}
\label{Paragraph:Implementation}
We implement \oursys with $\sim$10,000 lines of C++ code. 
We develop a KV Proxy as the KV execution engine based on code \cite{pingcapycsbgo} to support transaction semantics for LevelDB \cite{leveldb} and HBase \cite{hbase}. 
Since LevelDB is a standalone KV store, we implement data access interfaces by using brpc \cite{brpc} to enable remote data access.
OpenGauss \cite{opengauss} is a standalone database and only exposes its query interfaces for users. We modify its source code (less than 1,500 LoC), decouple the execution engines (openGauss-execution) and the storage engines (openGuass-MOT \cite{avni2020industrialmot}), and build an openGauss-execution - \oursys - openGauss-MOT database.
Similar to openGauss, we also modified NebulaGraph \cite{wu2022nebula} source code, enabling its execution engine (Nebula Graph) and storage engine (Nebula Storage) to connect to \oursys.

\Paragraph{\textbf{Physical Environment}}
\label{Paragraph:Physical Environment and Configurations}
Our experimental cluster is deployed on Aliyun, and each layer consists of 3 nodes by default. Each node (ecs.c6.4xlarge instance) is equipped with 16 vCPUs and 32 GB RAM and runs Ubuntu 22.04 LTS. All nodes are connected with a local area network of 5 Gbps.
The default epoch length of \oursys is 5ms. We use openGauss-execution - \oursys - openGauss-MOT database as our default choice.
Since openGauss-MOT \cite{avni2020industrialmot} is an in-memory store, we deployed it on a 16vCPU and 64GB RAM node to store a large amount of data in memory. 

\Paragraph{\textbf{Competitors}}
\label{Paragraph:Competitors}
We implement an epoch-based OCC protocol and a 2PL+2PC protocol under the same codebase with \oursys for comparison. The data is evenly divided according to the number of nodes. 
During the 2PC, the node, which receives the transaction requests will participate in the transaction committing as the coordinator.

To compare the performance in the cloud as well as across data models, we select TiDB \cite{huang2020tidb}, FoundationDB \cite{zhou2021foundationdb} and Epoxy \cite{kraft2023epoxy} as competitors.
TiDB is a compute-storage disaggregated NewSQL database with the storage layer handling concurrency control (CC). It uses optimistic execution and Percolator \cite{peng2010largepercolator} for transaction conflict resolution. 
FoundationDB (FDB) is a distributed key-value store that decouples CC from logging and storage, building a transaction-log-storage three-layer data store. 
FDB uses lock-free concurrency management with a deterministic transaction order and implements Serializable Snapshot Isolation (SSI) by combining OCC with MVCC. 
Epoxy is a middleware that enables connectivity to existing databases and ensures ACID transactions across heterogeneous storage systems by building a transaction metadata management layer. While Epoxy and CCaaS have different objectives in the architecture, both enhance distributed transaction processing. This shared approach makes Epoxy a relevant competitor for evaluating CCaaS’s performance in multi-engine and disaggregated databases.

\Paragraph{\textbf{Deployment}}
\label{Paragraph:Deployment}
We follow the official documents \cite{TiDBDeployment, FDBDeployment} to deploy TiDB and FoundationDB.
In TiDB, where the CC and storage layers are coupled, we deploy the execution layer with 3 machines, each with 16 vCPUs, and the storage layer with 3 machines, each with 32 vCPUs (16 for CC and 16 for storage).
For FDB, We deploy each layer with 3 machines, each with 16 vCPUs.
We deploy Epoxy with 3 machines, each with 16 vCPUs. Since the execution and data storage of the underlying databases (openGauss-MOT, LevelDB, Nebula) are coupled together, we deployed them on 32 vCPUs machines. 
We deployed stand-alone database instances on three nodes by default, forming a multi-master architecture.
Distributed databases (\eg HBase) are also built on a three-node configuration.
By default, we configure 3 client servers for each database, with each client-server hosting 32 clients that connect to the database and send transaction requests. When the number of execution nodes is expanded, the number of client servers scales correspondingly.

\begin{table}[t]
    \caption{Summary of Workloads}
    \Description[]{}
    \vspace{-0.15in}
    \label{tab:workloads}
    \centering
    \small
    {
	\setlength{\tabcolsep}{1.5mm}{
        \resizebox{0.48\textwidth}{!}{
        \begin{tabular}{cc|c|c}
		\hline
		\multicolumn{2}{c|}{\textbf{Name}} &
		{\textbf{Data Size}} &
		{\textbf{Operation Type}}
		\\
            \hline
		\multirow{2}{*}{\textbf{YCSB} \cite{YCSB}} & \multicolumn{1}{|c|}{{\textbf{YCSB-A}}}    & \multirow{2}{*}{1 M rows} & 10op/txn (50\% read - 50\% write)\\ 
            \cline{2-2} \cline{4-4}
		                                  & \multicolumn{1}{|c|}{{\textbf{YCSB-B}}}   &  & 10op/txn (95\% read - 5\% write) \\
		\hline
		\multicolumn{2}{c|}{\textbf{TPC-C} \cite{tpcc}}                    & 100 warehouse & 50\% NewOrder - 50\% Payment \\
		\hline
		\multicolumn{2}{c|}{\textbf{LDBC-SNB} \cite{LDBC-SNB}}                &  SF10 & 10op/txn (50\% fetch - 50\% insert)\\
		\hline
      \multicolumn{2}{c|}{\multirow{3}{*}{\textbf{Cross-Model}}}            & 1 M rows (KV)  & 10op/txn (5 KV, 4 SQL, 1 Graph)  \\
            &  & 1 M rows (SQL) & (KV, SQL: 80\% read - 20\% write) \\
            &  & SF10 (Graph) & (Graph: 50\% fetch - 50\% insert)\\
		\hline
	\end{tabular}
 }
        }
    }
    \normalsize
\end{table}

\Paragraph{\textbf{Workloads}}
\label{Paragraph:Workloads}
Table \ref{tab:workloads} presents the workloads used in the evaluation.
To make YCSB a transactional workload, we wrap 10 operations into transactions (txn) with a Zipfian access distribution.
For the TPC-C workload, we run a 50\% NewOrder - 50\% Payment mix to focuse on conflict resolution. In LDBC-SNB, we wrap 10 random fetch/insert operations into a transaction.
Additionally, to evaluate the cross-model transactions, we synthetically generate a multi-model transactional workload.


\subsection{Overall Performance}
\label{subsection:Performance}

\begin{figure}
    \centering
    {\includegraphics[width=0.48\textwidth]{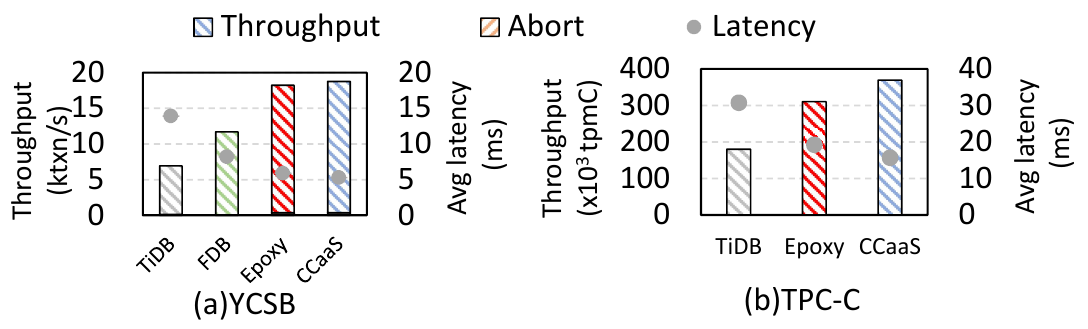}}
    \vspace{-0.3in}
    \caption{Experiment results with competitors.}
    \Description[]{}
    \label{fig:exp:performance}
\end{figure}

\begin{figure}
    \vspace{-0.17in}
    \centering
{\includegraphics[width=0.37\textwidth]{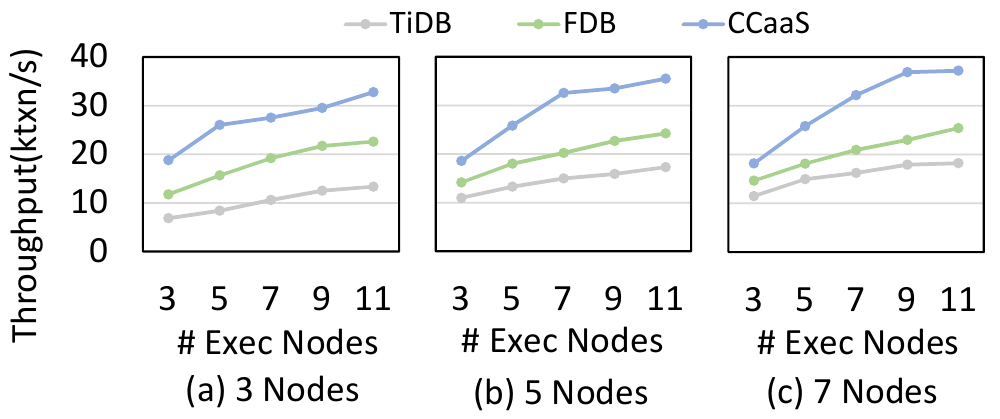}}
    \vspace{-0.17in}
    \caption{Comparison with existing disaggregated databases under YCSB-B workload.}
    \Description[]{}
    \label{fig:exp:performance-1}
\end{figure}

We compared \oursys with TiDB \cite{huang2020tidb}, FoundationDB \cite{zhou2021foundationdb} and Epoxy \cite{kraft2023epoxy}. Figure \ref{fig:exp:performance} reports the throughput and latency results. Since FoundationDB does not support TPC-C, we only compare TiDB and Epoxy for the TPC-C workload. For the YCSB-B workload, \oursys achieves 1.11-2.62 times higher throughput and 1.02-2.62 times lower latency than others. For the TPC-C workload, \oursys achieves 1.19-2.05 times higher throughput and 1.21-2.15 times lower latency. 
Additionally, we evaluate TiDB, FDB, and \oursys under the YCSB-B workload by varying the number of nodes per layer (from 3 to 7) and scaling the execution nodes while keeping other layers constant. 
The results in Figure \ref{fig:exp:performance-1} demonstrate that \oursys achieves 1.31–3.11 times higher throughput.

The performance of \oursys benefits from SM-OCC (Section \ref{subsection:Transaction Management}), allowing each node to resolve transaction conflicts independently. Specifically, our epoch-based mechanism only requires replicating write sets among the \oursys nodes, which reduces network round-trips (RTTs) caused by coordination. Additionally, the asynchronous log push-down method reduces transaction latency caused by updating data storage. The results can immediately be returned after finishing conflict resolution. In CCaaS, committing a transaction only requires 3 RTTs of network communication.

FoundationDB decouples logging from CC. Transaction commit results cannot be returned until the corresponding logs are replicated, requiring 4 RTTs and resulting in higher latency.
TiDB uses a variation of Percolator \cite{peng2010largepercolator} for CC and Raft \cite{ongaro2014searchraft} to replicate logs. Committing a transaction requires multiple RTTs (2PC + Raft, over 4.5 RTTs), increasing latency. Additionally, its pessimistic locking mechanism limits transaction concurrency. Epoxy uses multiple standalone databases for data storage. The execution time of a single statement in Epoxy is low. However, it needs to use Epoxy Coordinator and Epoxy Shim as proxies to forward user requests, maintain additional meta-information to ensure MVCC across multiple storage engines, and use an optimized two-phase commit protocol to ensure that transactions are committed atomically, introducing a certain amount of latency into the system. Therefore, its performance is approximate to that of CCaaS using SM-OCC for concurrency control.

\begin{figure*} [t]
\centering
    \begin{minipage}{0.45\textwidth}
    \centering
    \subfloat[YCSB-A]{\includegraphics[width=0.45\linewidth]{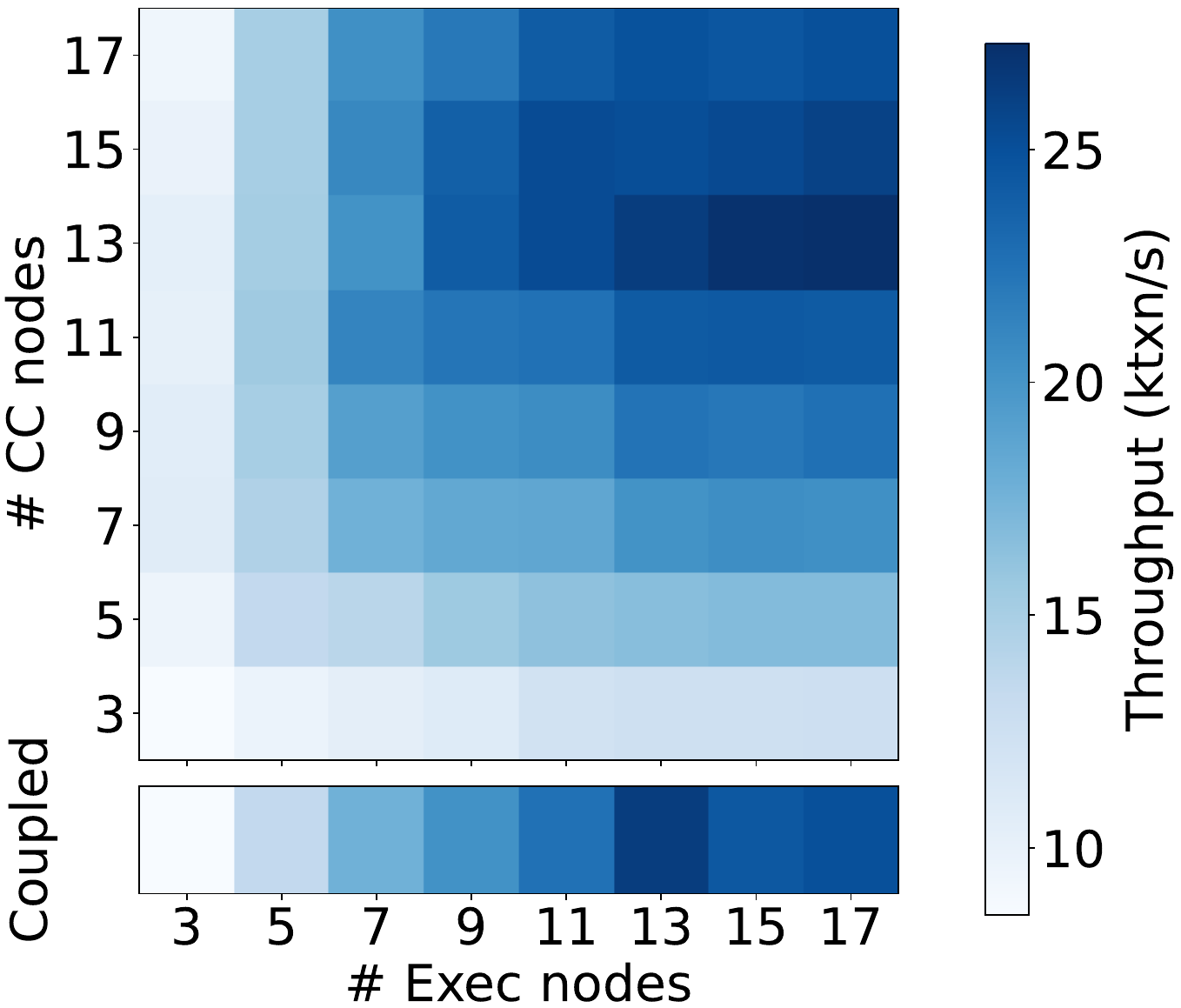}}
    \hspace{0.05in}
    \subfloat[YCSB-B]{\includegraphics[width=0.45\linewidth]{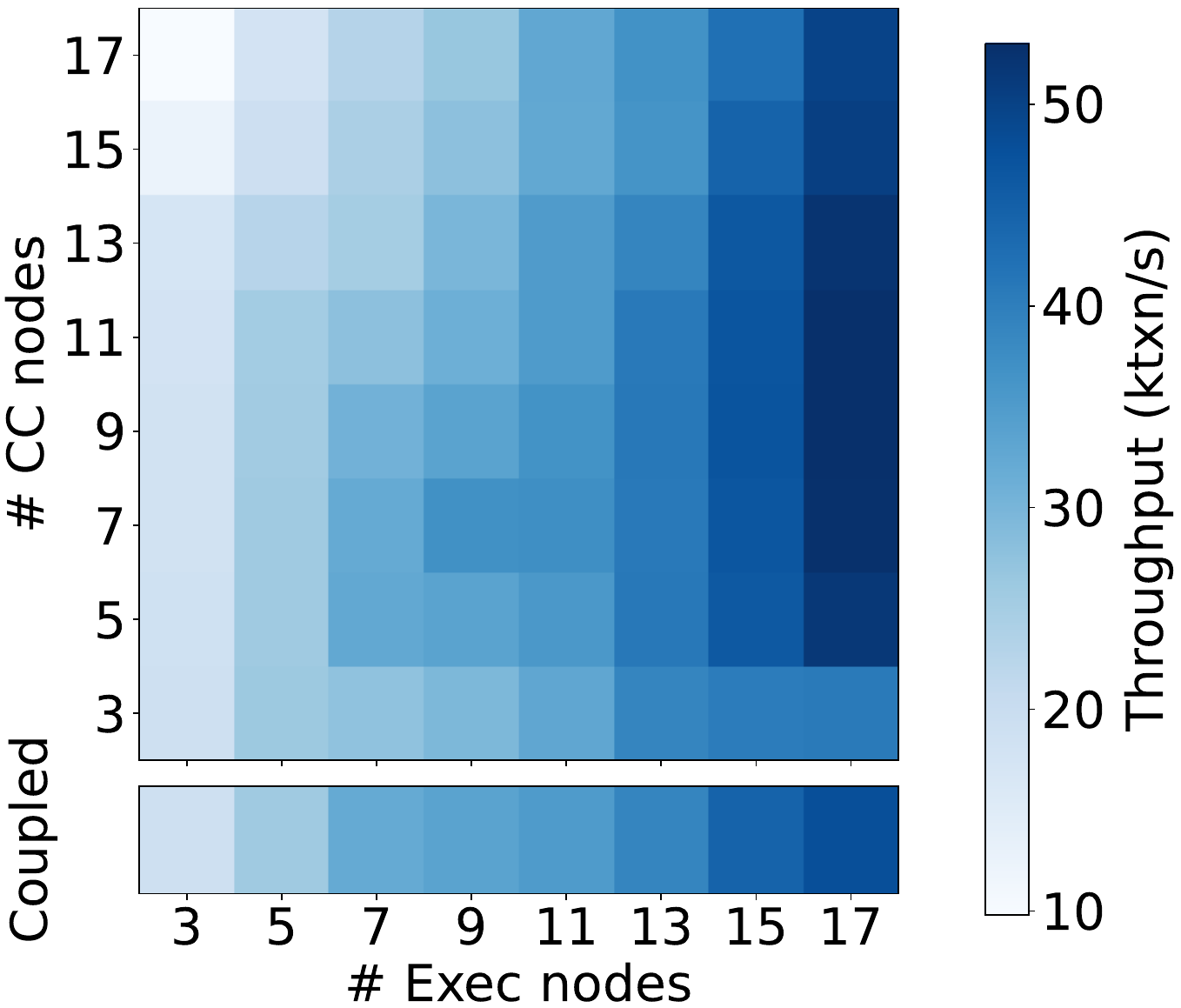}}
    \vspace{-0.15in}
    \caption{Throughput when scaling the number of execution nodes and CC nodes in CCaaS.}
    \label{fig:exp:Scalability}
    \end{minipage}
    \hspace{0.5in}
    \begin{minipage}{0.42\textwidth}
    \centering
    {\includegraphics[width=1\linewidth]{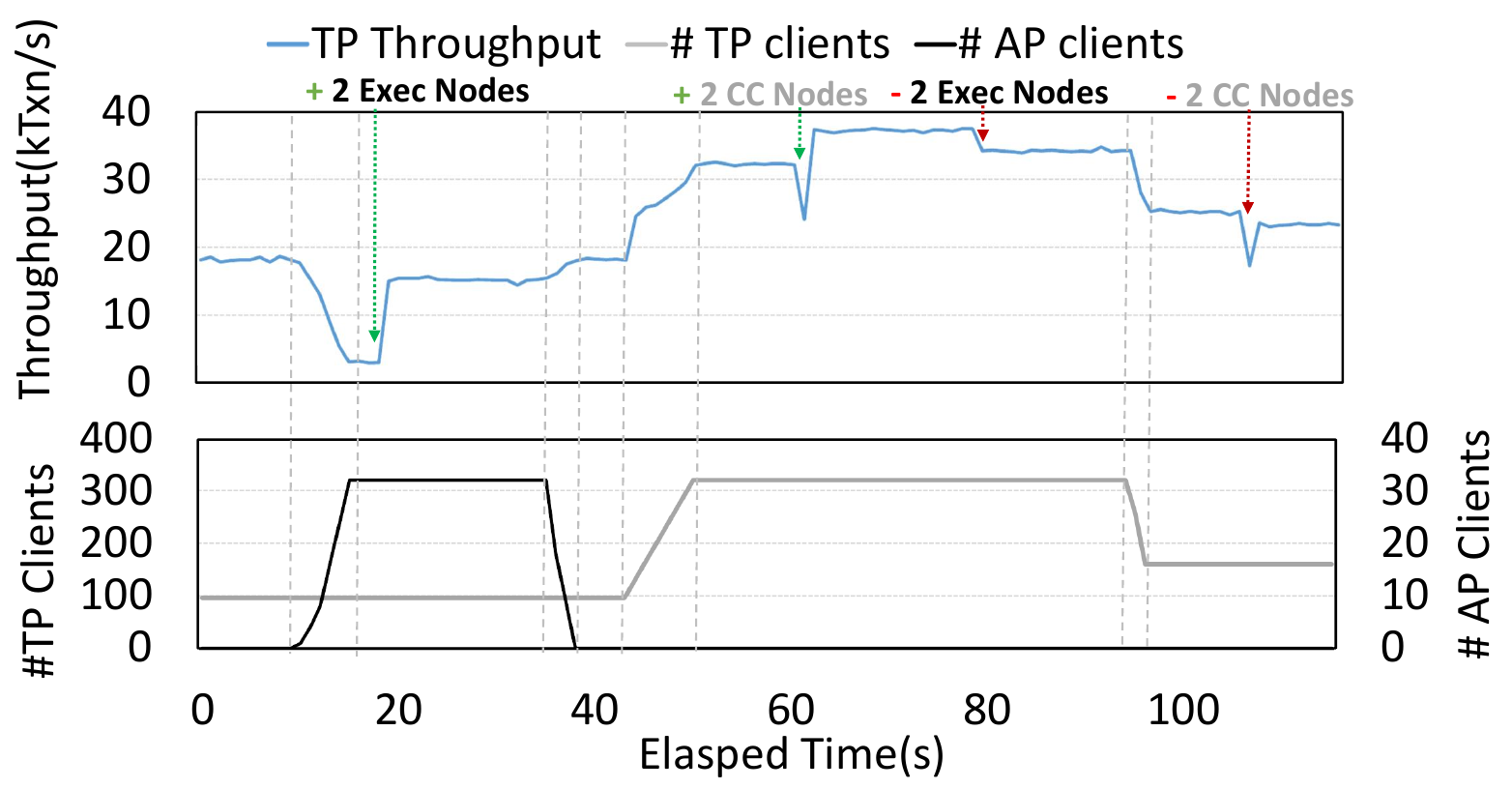}}
    \vspace{-0.3in}
    \caption{Elasticity performance under changing workloads.}
    \Description[]{}
    \label{fig:exp:Elasticity}
    \end{minipage}
    \vspace{-0.1in}
\end{figure*}

\begin{figure*} [t]
\centering
    \begin{minipage}{0.43\textwidth}
        \centering
        \centerline{\includegraphics[width=1\linewidth]{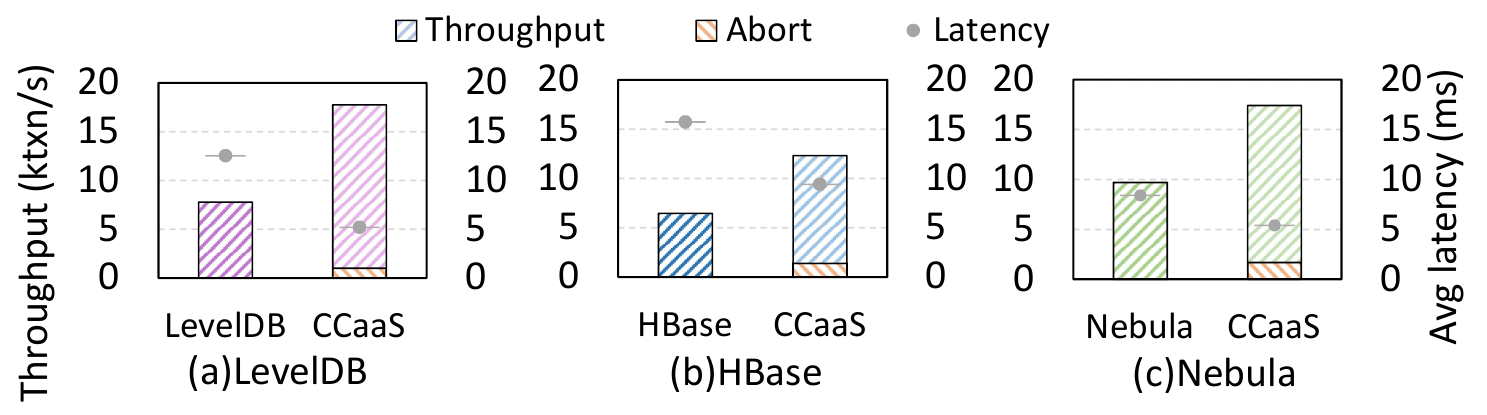}}
        \vspace{-0.1in}
        \caption{Performance of NoSQL DBs empowered with ACID TP capability.}
        \Description[]{}
        \label{fig:exp:ap}
    \end{minipage}%
    \hspace{0.1in}
    \begin{minipage}{0.25\textwidth}
        \centering
        \centerline{\includegraphics[width=1\linewidth]{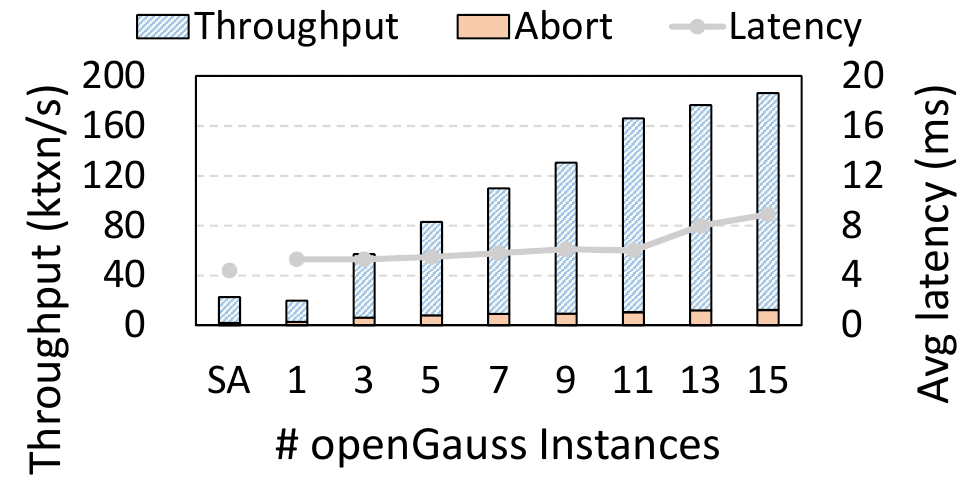}}
        \vspace{-0.15in}
        \caption{Performance of serving standalone TP engines.}
        \Description[]{}
        \label{fig:exp:tp}
    \end{minipage}%
    \hspace{0.1in}
    \begin{minipage}{0.27\textwidth}
        \centering
        \includegraphics[width=1\linewidth]{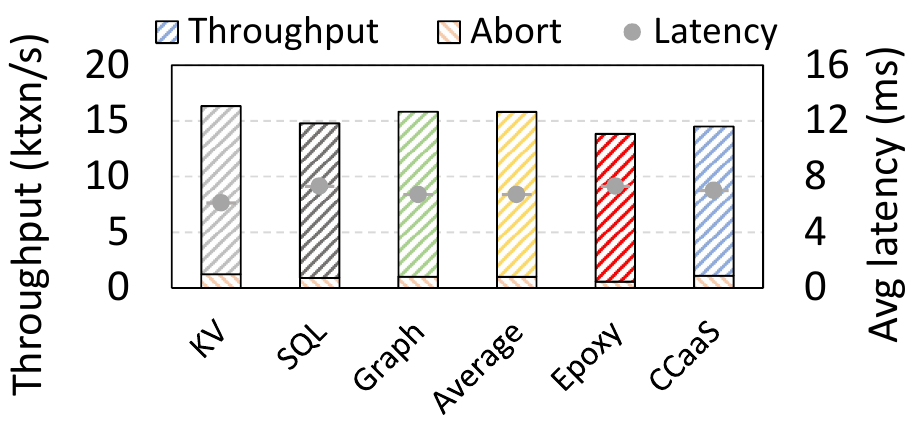}
        \vspace{-0.25in}
        \caption{Performance of cross-model (CM) transactions.}
        \Description[]{}
        \label{fig:exp:multimodel}
    \end{minipage}%
    \vspace{-0.1in}
\end{figure*}

\subsection{Scalability}
\label{subsection:Scalability}
In this experiment, we evaluate the scaling performance when CC is coupled with the execution engine and when scaling execution nodes and CC nodes independently.
The results in Figure \ref{fig:exp:Scalability} show that the CC-execution decoupled system shows better throughput performance than the CC-execution coupled system, with various CC-execution node number combinations under both YCSB-A and YCSB-B workloads. Especially under the YCSB-B workload, the optimal performance is obtained with 17 execution nodes and 7 CC nodes. 
This verifies the necessity of independent scaling of CC, which also requires much less cost.
In addition, CCaaS shows better performance under the read-intensive YCSB-B workload than under the write-intensive YCSB-A workload. This is because large write sets incur more network overhead in the CC-decoupled system, reducing system performance.

\subsection{Elasticity}
\label{subsection:elasticity}
To evaluate the elasticity improvements facilitated by CCaaS, we test the system under dynamic workloads using a mix of TPC-H and YCSB-B. TPC-H represents computationally heavy analytical processing (AP) workloads handled by the execution layer, while YCSB-B models transactional processing (TP) workloads requiring conflict resolution in the CC layer. 
We dynamically adjust the number of AP and TP clients over time to simulate the dynamic AP-TP mixed workloads, as shown at the bottom of Figure \ref{fig:exp:Elasticity}. In response to these workload changes, we dynamically adjust the number of nodes in the execution and CC layers and observe changes in TP throughput, as shown in the upper part of Figure \ref{fig:exp:Elasticity}. 

As shown in Figure \ref{fig:exp:Elasticity}, when the AP workload increases at 12s, TP throughput drops due to CPU contention in the execution layer. To mitigate this, we scale out the execution layer by adding 2 nodes at 18s, quickly restoring TP throughput.
At 45s, increasing TP clients makes the CC layer a bottleneck due to its limited capacity for handling concurrent transactions. Scaling out the CC layer at 60s improves throughput but triggers re-sharding in CCaaS, causing a temporary 27\% drop due to the redistribution of committed transaction metadata.
After 40s, the AP workload ends, reducing computational demand. At 80s, we remove 2 execution nodes to save costs, causing a slight TP throughput drop. At 100s, with fewer TP clients, the CC layer is over-provisioned, so we remove 2 CC nodes, with minimal performance impact.
The results demonstrate that the decoupled CC layer enables flexible resource allocation and efficient response to workload changes. By allowing independent scaling of the execution and CC layers, 
CCaaS enhances elasticity while reducing costs.

Notably, CCaaS supports re-sharding without interrupting the CC service. 
When re-sharding begins, CCaaS first requires that conflict resolution follows the new sharding policy in a pre-determined number of epochs. During re-sharding, CC nodes transfer metadata (snapshot and subsequent metadata updates) to designated nodes while continuing conflict resolution under the original sharding strategy. Once re-sharding is completed, CCaaS switches to the new sharding policy and resumes normal processing.

\begin{figure*} [th]
\centering
    \begin{minipage}{0.24\textwidth}
        \begin{center}
        \vspace{0.08in}
        \includegraphics[width=1\linewidth]{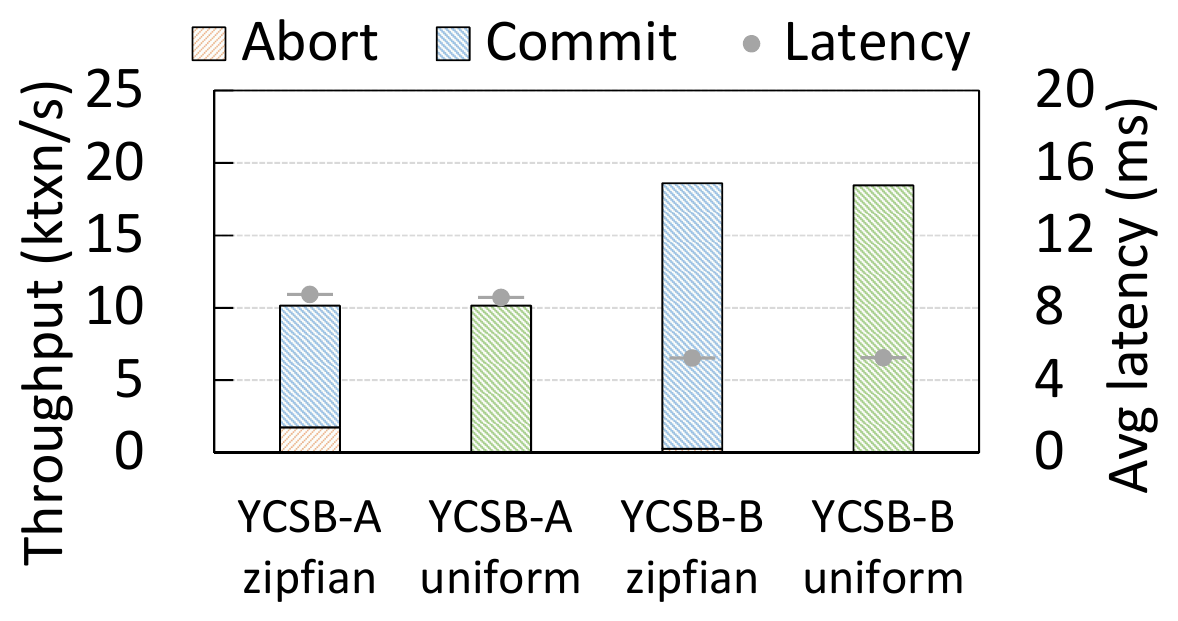}
        \vspace{-0.25in}
        \caption{Performance under different contention.}
        \Description[]{}
        \label{fig:exp:contention}
        \end{center}
    \end{minipage}%
    \hspace{0.05in}
    \begin{minipage}{0.255\textwidth}
        \begin{center}
        \includegraphics[width=1\linewidth]{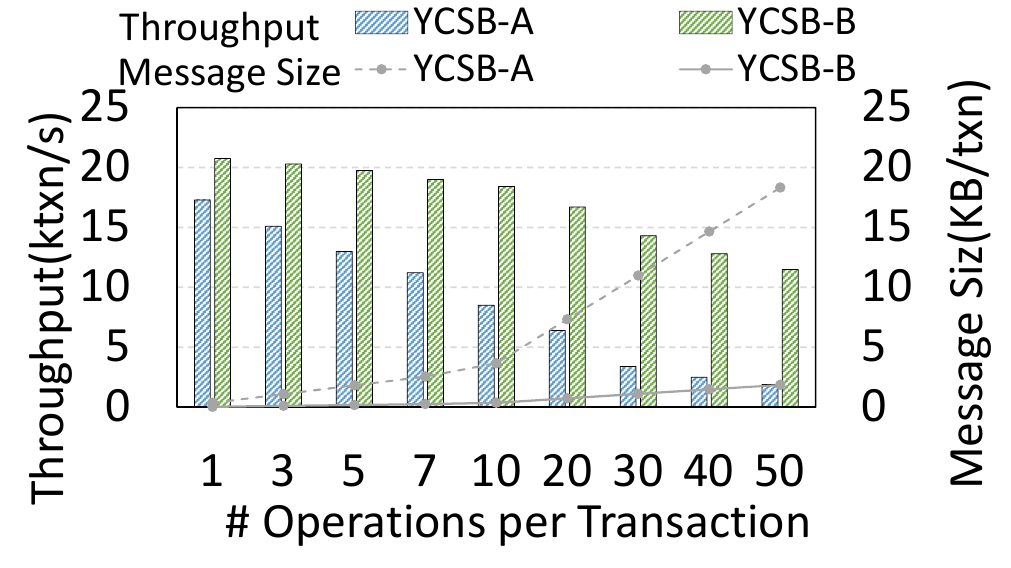}
        \vspace{-0.3in}
        \caption{Performance when varying number of operations.}
        \Description[]{}
        \label{fig:exp:datasize}
        \end{center}
    \end{minipage}%
    \hspace{0.05in}
    \begin{minipage}{0.23\textwidth}
        \begin{center}
        \vspace{0.2in}
        \includegraphics[width=1\linewidth]{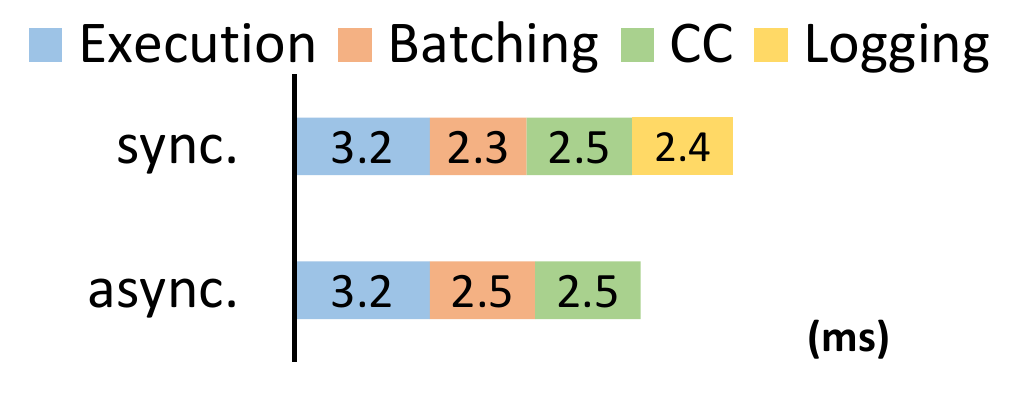}
        \caption{Runtime breakdown (sync./async. logging).}
        \Description[]{}
        \label{fig:exp:runtime}
        \end{center}
    \end{minipage}%
    \hspace{0.05in}
    \begin{minipage}{0.24\textwidth}
        \begin{center}
        \includegraphics[width=1\linewidth]{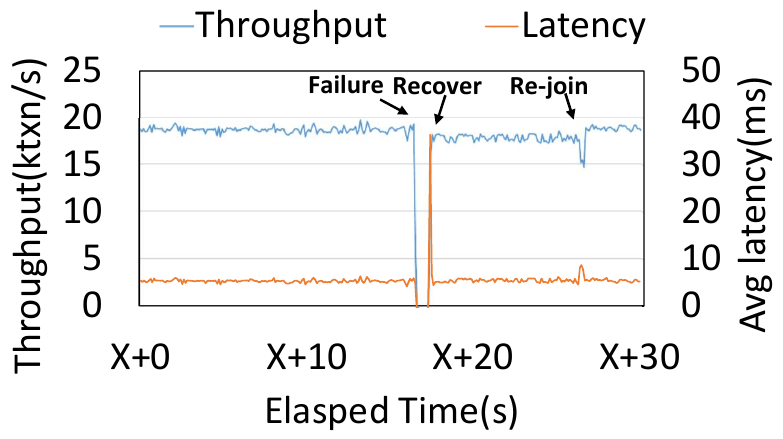}
        \vspace{-0.28in}
        \caption{Performance variation when system recovery.}
        \Description[]{}
        \label{fig:exp:fault}
        \end{center}
    \end{minipage}%
    \vspace{-0.1in}
\end{figure*}

\subsection{Case Studies}

\Paragraph{{Supporting TP for NoSQL DBs}}
To demonstrate the TP performance provided by \oursys, we use the YCSB-B workload to evaluate the throughput and the average latency of original NoSQL databases and that of the \oursys-enhanced ones.
As shown in Figure \ref{fig:exp:ap}, by connecting to \oursys, these NoSQL databases gain transaction processing capability and show higher operation throughput and lower latency due to the log asynchronous push-down method. 

\Paragraph{{Building a Multi-Master Database}}
Figure \ref{fig:exp:tp} shows the TP’s horizontal scalability by connecting multiple openGauss instances to \oursys under the YCSB-B workload.
When only one openGauss instance is connected to \oursys, the performance is lower, and latency is higher compared to the standalone instance due to the additional network I/O introduced by \oursys. Decoupling allows modules to scale independently to meet resource demands under varying workloads. However, when the workload is within the capacity of a single instance, the benefits of decoupling are not realized, and the network overhead becomes a burden.
In contrast, as system load increases, adding more execution nodes allows the workload to be distributed across multiple nodes, avoiding single-point bottlenecks and improving throughput. 

\Paragraph{{Cross-Model Transactions}}
We use the cross-model workload to test \oursys and Epoxy. 
For comparison, we use the same data to generate a single model transaction (10 op/txn) to test each single-model storage engine. The weighted average throughput and latency of these single-model transactions are also reported. The results are shown in Figure \ref{fig:exp:multimodel}. 
As shown, the KV storage engine is the fastest since it does not incur overhead from statement parsing, execution plan generation, \etc
Due to the processing of Graphs in the workload is simpler than that of SQL, SQL has the lowest performance with the highest latency. Because \oursys needs to collect all single-model sub-transactions, the performance of multi-model transactions is determined by the slowest engine.
In supporting for cross-model transaction processing, Epoxy demonstrates performance that approximates that of CCaaS.

\subsection{\textbf{Varying Workloads}}

We evaluate the performance of \oursys under zipfian and uniform access distributions using YCSB workloads. As shown in Figure \ref{fig:exp:contention}, a high write rate with Zipfian distribution will cause more transactions to be aborted. This is because the SM-OCC only commits one transaction when multiple transactions update the same record. Under the Zipfian distribution, transactions are easier to access the same records, causing a higher abort rate. 
Comparing system performance under YCSB-A and B workloads, \oursys has better performance under YCSB-B workloads. This is because under YCSB-A load, transactions have more write operations, causing higher network overhead in \oursys and reducing system performance.

Additionally, we adjust the operation count of transactions to evaluate system performance with varying read-write set sizes. Figure \ref{fig:exp:datasize} shows that the system works well with low operation numbers. Adding more operations lead to longer execution times and larger message sizes, reducing throughput. \oursys performs better under the read-intensive workload (YCSB-B) because it validates reads by checking metadata in the read set (\eg CSN), which increases slowly. In the write-intensive workload (YCSB-A), additional write operations significantly increase the message size, and increasing write operations increases transmission and replication overhead in \oursys. The write sets replication consume a lot of computation resources, causing performance degradation.

\subsection{\textbf{Sync. Logging vs. Async. Logging}}
As shown in Figure \ref{fig:exp:runtime}, we perform a breakdown analysis for a single transaction under the YCSB-A workload. 
Since YCSB-A is a write-intensive workload, the log push-down takes 2.4 ms on average per transaction. In comparison, the log push down in the read-intensive workload YCSB-B is less than 1ms (not shown in Figure \ref{fig:exp:runtime} due to space constraints). When using asynchronous logging, the overall latency is reduced. However, since more transactions are processed per epoch, the batching and CC phases take longer.

\subsection{\textbf{Fault Recovery}}
To evaluate performance under failure, we manually shut down a \oursys node and see how \oursys acts. Figure \ref{fig:exp:fault} shows the changes in throughput and latency. There is a temporary performance drop following a node failure at 16 seconds, as active nodes wait for write sets from the failed node. \oursys quickly responds to this failure due to Raft-based membership management (with a 500 ms timeout setup). 
After changing the leader of the Raft instances (Section \ref{subsection:Fault Recovery}), \oursys resumes providing service. With only two nodes in \oursys, overall throughput slightly decreases and latency increases due to the increased load on each node. After the crashed node recovers (at 26 seconds), the Raft-based membership management notices and adds the recovery node back to the cluster, and then the throughput and latency return to normal.

\section{Related Work}
\label{section:7}
\label{section:Related Work}

\Paragraph{Decouple Transaction Management Component}
Earlier works \cite{lomet2009lockingkeyranges, lomet2009unbundlingtransactionservicesinthecloud, levandoski2011deuteronomy, shacham2018takingomid, das2013elastras, eldeeb2016transactions} introduce transaction components (TM) to handle conflicts via virtual resources. Deuteronomy \cite{levandoski2011deuteronomy} uses TM to provide transaction processing (TP) capabilities for KV stores. In \cite{eldeeb2016transactions}, TM is used to detect transaction dependencies and release locks early during 2PC. Omid \cite{ferro2014omid, bortnikov2017omid, shacham2018takingomid} and Tell \cite{loesing2015designtell} use TM to implement multi-version concurrency control (MVCC), improving transaction throughput. DIBS \cite{gaffney2021dibs} implements the TM with predicate locking to guarantee transaction isolation. These systems generally focus on enabling TP for existing data stores, using centralized TM to resolve conflicts. FoundationDB \cite{zhou2021foundationdb} decouple the TM based on the roles (coordinator and participant) in the 2PC for better scalability. \oursys follows the principle of cloud-native design to decouple the CC layer, tries to use a multi-master architecture to improve the scalability of CC, and supports processing with multiple models by abstracting the CC from the data models.

\Paragraph{Cross-Engine Transaction}
Conventionally, cross-data store transactions are implemented through a distributed transaction protocol such as X/Open XA \cite{specification1991theXASpecification} or WS-TX \cite{WSAtomicTransaction}. Such protocols use two-phase commit to ensure atomicity. 
Cherry Garcia \cite{dey2015scalableCherryGarcia} and Omid \cite{ferro2014omid} provide ACID transactions across multiple key-value stores but only support key-value operations.
Skeena \cite{zhang2022skeena} proposes a holistic approach to cross-engine
transactions and uses an atomic commit protocol to efficiently ensure correctness
and isolation. Epoxy \cite{kraft2023epoxy} provides ACID transactions across heterogeneous data stores by using an additional MVCC control panel, but it needs a primary transactional DBMS as a transaction coordinator for transaction processing.
In comparison, \oursys makes the CC module an independent service, allowing it to be connected by various data stores concurrently.
By maintaining transaction meta-information and resolving conflicts at epoch-granularity, \oursys allows the system to connect to multiple storage engines simultaneously.

\section{Conclusion}
\label{section:8}
\label{sec:conclusion}
This paper proposes Concurrency Control as a Service (\oursys), an execution-CC-storage three-layer database architecture. 
We demonstrate that databases can be revolutionized with \oursys: NoSQL databases can gain ACID TP ability, standalone TP databases can support distributed transactions with horizontal scalability, and cross-model transactions can be realized. 
Our evaluation results show that \oursys outperforms existing disaggregated databases, including TiDB and FoundationDB, and exhibits a certain degree of scalability of transaction processing. 
CC, as an independent service, has potential that has not been fully developed. Our future work will focus on optimizations like the support of more isolation levels and consistency choices, and even cross-device CC service. 

\begin{acks}
This work was supported by the National Key R\&D Program of China (2023YFB4503601), the National Natural Science Foundation of China (62461146205), the Distinguished Youth Foundation of Liaoning Province (2024021148-JH3/501), and Huawei. Yanfeng Zhang and Zeshun Peng are the corresponding authors.
\end{acks}



\bibliographystyle{ACM-Reference-Format}
\bibliography{all}

\end{document}